\newcommand{\Draftmode}{\False}

\newcommand{\arxiv}{\True}  % make arxiv version

\newcommand{\version}{\Ifthenelse{\arxiv}{Arxiv}{CAV} version}

\documentclass[runningheads]{llncs}

\usepackage{latexsym}
\usepackage{amssymb}
\usepackage{amsmath}
\usepackage{graphicx}
\usepackage{xcolor}
\usepackage{ifthen}
\usepackage{listings}
\usepackage{dsfont}
\usepackage[noend]{algorithmic}
\usepackage{algorithm}
\usepackage{tabularx}
\usepackage{rotating}
\usepackage{tikz}
\usepackage{enumitem}
\usepackage{cleveref}
\usepackage{url}

\tikzset{every picture/.style={remember picture}}

\definecolor{darkgreen}{RGB}{0,100,0}

\newcommand{\localCommand}{} % command names start with \
\newcommand{\newlengthsettowidth}[2]{\newlength {#1} \settowidth{#1}{#2}}
\newcommand{\newcounterset}      [2]{\newcounter{#1} \setcounter{#1}{#2}}

\newcommand{\numberRange}[1]{\mathbb #1}
\newcommand{\NN}{\numberRange N}                        % naturals
\newcommand{\RR}{\numberRange R}                        % reals
\newcommand{\range}[3][X]{\Ifthen{\Equal{#1}{X}}{\{}#2,\ldots,#3\Ifthen{\Equal{#1}{X}}{\}}} % {1,...,n}, or 1,...,n if no optional arg
\newcommand{\atl}{\geq}                                 % at least
\newcommand{\atm}{\leq}                                 % at most
\newcommand{\union}       {\mathbin        {\cup}}      % set union
\newcommand{\Union}       {\operatorname{\bigcup}}      % big union
\newcommand{\Sum}         {\sum}                        % (for uniformity)
\newcommand{\Product}     {\prod}                       % (for uniformity)

\newcommand{\superseteq} {\supseteq}                    % superset ">="
\newcommand{\func}[3]{#1 \colon #2 \rightarrow #3}      % function f:A->B
\newcommand{\ceils}   [1]{     \lceil  #1       \rceil } % argument in ceilings
\newcommand{\limplies}{\Rightarrow}               % logical 'implies'
\newcommand{\true}    {\mathit{true}}

\newcommand{\Ifthenelse}[3]{\ifthenelse{#1}{#2}{#3}}   % for consistency: initial capital
\newcommand{\Ifthen}    [2]{\Ifthenelse{#1}{#2}{}}
\newcommand{\Unless}    [2]{\Ifthen{\not {#1}}{#2}}
\newcommand{\Equal}     [2]{\equal{#1}{#2}}            % for consistency: initial capital
\newcommand{\Empty}     [1]{\Equal{#1}{}}
\newcommand{\True}         {\Equal{1}{1}}
\newcommand{\False}        {\Equal{1}{2}}

\providecommand{\Draftmode}{\True} % on by default, but you can turn it off on the latex command line

\newcommand{\Itedraft}    [2]{\Ifthenelse{\Draftmode}{#1}{#2}}
\newcommand{\Ifdraft}     [1]{\Itedraft{#1}{}}

\newcommand{\drafttext}   [2][\draftcolor]{\Ifdraft{{\color{#1}#2}}}
\newcommand{\draftpointer}{\makebox[0mm][c]{$^{*}$}}
\newcommand{\draftmargin} [2][\draftcolor]{\drafttext[#1]{\draftpointer\marginpar[\raggedleft\small{\color{#1}#2}]{\raggedright\small{\color{#1}#2}}}} % marginalia, shown in draftmode only

\newcommand{\draftnewpage}  {\Ifdraft{\newpage}}

\newcommand{\draftappendix}[1][Draft Appendix]{\Itedraft{\clearpage \section*{#1}}{\end{document}}}

\newcommand{\draftauthor}[3]{
  \newcommand{#1}[1]{\drafttext  [#3]{##1}}  % Use:     \draftauthor{\thomas}{\thomasmargin}{green}
  \newcommand{#2}[1]{\draftmargin[#3]{##1}}  % defines: \thomas{long comment}, \thomasmargin{short comment}
}

\newcommand{\Ifspacecolor}{gray}
\draftauthor{\Ifspace}{\Ifspacemargin}{\Ifspacecolor}

\Ifdraft{\setlength{\overfullrule}{10mm}}

\newcommand{\mboxscript}[1]{\mbox{\scriptsize#1}}

\newcommand{\centerOne}     [1]{\mbox{} \hfill #1 \hfill \mbox{}}                     % |    #1    |. Use with  one        object.
\newcommand{\centerTwoIn}   [2]{\mbox{} \hfill #1 \hfill #2 \hfill \mbox{}}           % |  #1  #2  |. Use with  two  small objects.
\newcommand{\centerTwoOut}  [2]               {#1 \hfill #2}                          % |#1      #2|. Use with  two  large objects.
\newlength{\posLength}

\newcommand{\mc}{\multicolumn}                          % abbreviation for 'multicolumn'
\newlengthsettowidth{\tabLength}{\ \ \ }
\newcommand{\wbox}[2][\tabLength]{\hspace*{#1}\mbox{#2}\hspace*{#1}}

\newcommand{\Paragraph}[1][]{\vspace{\baselineskip}\Unless{\Empty{#1}}{\noindent\emph{#1}\ }} % no empty lines after \Paragraph{Header.}, otherwise the '\noindent' is void
\newcommand{\End}{\end{document}}                       % ignore rest of document
\newcommand{\format}{\drafttext{\underline{\makebox[0mm][c]{$\mid$}}}} % Mark commands for final formatting: \format\vspace{2mm}. Also creates visible mark (of zero width) in text
\newcommand{\horiline}[1][\textwidth]{\noindent\rule{#1}{1pt}}   % horizontal line to separate text portions. Use on a separate line
\newcommand{\emphasize}[1]{\textbf{#1}}                 % stronger than \emph
\newcommand{\emphdef}[1]{\emphasize{#1}}                % \emph is too weak in defs

\newcommand{\namecolor}[2][]{\Ifthenelse{\Empty{#1}}{{\color{#2}#2}}{{\color{#2}#1}}}

\newcommand{\defFullOrAbbrev}[2]{#1} % full name: Definition/definition
\newtheorem{ASS}     {\defFullOrAbbrev{Assumption} {Assn.}}

\newtheorem{COR}[ASS]{\defFullOrAbbrev{Corollary}  {Cor.}}
\newtheorem{DEF}[ASS]{\defFullOrAbbrev{Definition} {Def.}}
\newtheorem{EXA}[ASS]{\defFullOrAbbrev{Example}    {Ex.}}
\newtheorem{LEM}[ASS]{\defFullOrAbbrev{Lemma}      {Lem.}}

\newtheorem{PRO}[ASS]{\defFullOrAbbrev{Property}   {Prop.}}

\newtheorem{THE}[ASS]{\defFullOrAbbrev{Theorem}    {Thm.}}

\newcommand{\opNote}[2]{\stackrel{\mbox{\tiny #1}}{#2}}

\newlength{\spaceAfterEOP}
\newcommand{\Proof}{\noindent\textbf{Proof}}               % introducing a proof
\newcommand{\eop}[1][\spaceAfterEOP]{\eopBox \vspace{#1}}  % end of proof (box), followed by new paragraph. Use:
\newcommand{\eopBox}{~\hfill$\Box$}                        % box at end of proof. Use only where new paragraph
   \renewcommand{\defFullOrAbbrev}[2]{#2}
\newcommand{\refCapitalOrSmall}[3]{#1#3} % always capitalized: Definition/Def.
\newcommand{\refFullOrAbbrev}[2]{#1}     % full name: Definition/definition
\newcommand{\algorithmref}  [2][!]{\genericref[#1] A a {\refFullOrAbbrev{lgorithm}  {lg.}}  {algorithm}  {#2}}
\newcommand{\appendixref}   [2][!]{\genericref[#1] A a {\refFullOrAbbrev{ppendix}   {pp.}}  {appendix}   {#2}}

\newcommand{\definitionref} [2][!]{\genericref[#1] D d {\refFullOrAbbrev{efinition} {ef.}}  {definition} {#2}}
\newcommand{\exampleref}    [2][!]{\genericref[#1] E e {\refFullOrAbbrev{xample}    {x.}}   {example}    {#2}}
\newcommand{\figureref}     [2][!]{\genericref[#1] F f {\refFullOrAbbrev{igure}     {ig.}}  {figure}     {#2}}
\newcommand{\itemref}       [2][!]{\genericref[#1] I i {\refFullOrAbbrev{tem}       {tem}}  {item}       {#2}}
\newcommand{\lemmaref}      [2][!]{\genericref[#1] L l {\refFullOrAbbrev{emma}      {em.}}  {lemma}      {#2}}
\newcommand{\lineref}       [2][!]{\genericref[#1] L l {\refFullOrAbbrev{ine}       {ine}}  {line}       {#2}}

\newcommand{\propertyref}   [2][!]{\genericref[#1] P p {\refFullOrAbbrev{roperty}   {rop.}} {property}   {#2}}

\newcommand{\sectionref}    [2][!]{\genericref[#1] S s {\refFullOrAbbrev{ection}    {ec.}}  {section}    {#2}}
\newcommand{\tableref}      [2][!]{\genericref[#1] T t {\refFullOrAbbrev{able}      {able}} {table}      {#2}}
\newcommand{\theoremref}    [2][!]{\genericref[#1] T t {\refFullOrAbbrev{heorem}    {hm.}}  {theorem}    {#2}}

\newcommand{\equationref}[2][!]{\Ifthen{\Equal{#1}!}{\refCapitalOrSmall E e {\refFullOrAbbrev{quation}{q.}}~}(\ref{equation: #2})}

\newcommand{\genericref} [6][!]{\Ifthen{\Equal{#1}!}{\refCapitalOrSmall{#2}{#3}{#4}~}\ref{#5: #6}} % args: [Append.~]{A}{a}{ppendix}{appendix}{Proofs}
 \renewcommand{\refFullOrAbbrev}[2]{#2}
\algsetup{indent=1.3em}

\newcommand{\plkeyword}[1]{\textbf{#1}}

\renewcommand{\algorithmicrequire}{\plkeyword  {Input}:}

\newcommand{\plgoto}   {\plkeyword{goto}}

\newcommand{\plif}     {\plkeyword{if}}

\newcommand{\plrepeat} {\plkeyword{repeat}}

\newcommand{\plfor}    {\plkeyword{for}}

\newcommand{\plreturn} {\plkeyword{return}}
\newcommand{\plassert} {\plkeyword{assert}}

\newcommand{\code}[1]{\texttt{#1}}

\draftauthor{\andrew}{\andrewmargin}{blue}
\draftauthor{\thomas}{\thomasmargin}{green}

\renewcommand{\Im}{\mathit{Im}}
\newcommand{\pc}  {\mathit{pc}}

\newcommand{\concStates} {\mathcal S}
\newcommand{\abstrStates}{\mathcal A}
\newcommand{\absIm}     {\overline{\Im}}

\newcommand{\parameterize}[2]{{#1}^{#2}}    % superscript form
\renewcommand{\parameterize}[2]{{#1}({#2})} % parenthetical form. This seems to take more space
\newcommand{\parameterizeUnlessEmpty}[2]{\Ifthenelse{\Empty{#2}}{#1}{\parameterize{#1}{#2}}}
\renewcommand{\RR}        [1][]{\parameterizeUnlessEmpty{\mathit{RR}}        {#1}} % Round Robin scheduler
\newcommand  {\reached}   [1][]{\parameterizeUnlessEmpty{R}                  {#1}}
\newcommand  {\absreached}[1][]{\parameterizeUnlessEmpty{\overline{\reached}}{#1}}

\newcommand{\prg}     {\mathcal P}
\newcommand{\otherprg}{\mathbb  P}
\newcommand{\Bprg}    {\mathbb  B}

\newcommand{\prop} C % Correctness ('P' is taken)

\newcommand{\FinishRounds}{\mathit{FinishRounds}}
\newcommand{\Unexplored}  {\mathit{Unexplored}}
\newcommand{\Reached}     {\mathit{Reached}}
\newcommand{\Frontier}    {\mathit{Frontier}}
\newcommand{\finder}      {\mathit{finder}}
\newcommand{\roundsTaken} {\mathit{rounds\_taken}}
\newcommand{\delaysTaken} {\mathit{delays\_taken}}
\newcommand{\Image}       {\mathit{Image}}

\newcommand{\step}[4]{#1 \stackrel{\mboxscript{$#2:#3$}}{\longrightarrow} #4}

\newcommand{\wqo}{\preccurlyeq}

\newcommand{\duba}{DrUBA}

\newcommand{\ch}[1]{\multicolumn 1 c {{\scriptsize #1}}}
\newcommand{\Time} {\multicolumn 1 {c|} {Time:}}

\Ifthenelse{\arxiv}{%
  \newcommand{\etr}{\\ (Extended Technical Report)}}{%
  \newcommand{\etr}{}}

\title{%
  Delay-Bounded Scheduling Without Delay!\etr\thanks{Partially supported by the US National Science Foundation under grant \#1718235}}

\author{%
  Andrew Johnson\inst{1} \and
  Thomas Wahl\inst{1,2}}

\institute{%
  Northeastern University, Boston MA, 02115, USA \and
  GrammaTech Inc.}

\begin{document}

\maketitle

\begin{abstract}
  We consider the broad problem of analyzing safety properties of
  asynchronous concurrent programs under arbitrary thread
  interleavings. \emph{Delay-bounded deterministic scheduling}, introduced
  in prior work, is an efficient bug-finding technique to curb the large
  cost associated with full scheduling nondeterminism. In this paper we
  first present a technique to \emph{lift the delay bound} for the case of
  finite-domain variable programs, thus adding to the efficiency of bug
  detection the ability to prove safety of programs under arbitrary thread
  interleavings. Second, we demonstrate how, combined with predicate
  abstraction, our technique can both refute and verify safety properties
  of programs with unbounded variable domains, even for unbounded thread
  counts. Previous work has established that, for non-trivial concurrency
  routines, predicate abstraction induces a highly complex abstract program
  semantics.
% that goes beyond even the rich class of well-quasiordered systems.
  Our technique, however, never statically constructs an abstract
  parametric program; it only requires some abstract-states set to be
  closed under certain actions, thus eliminating the dependence on
  the existence of verification algorithms for
% (e.g., well-quasiordered)
  abstract programs. We demonstrate the efficiency of our technique on many
  examples used in prior work, and showcase its simplicity compared to
  earlier approaches on the unbounded-thread Ticket Lock protocol.
\end{abstract}

\draftnewpage

\section{Introduction}

Asynchronous concurrent programs consist of a number of threads executing
in an interleaved fashion and communicating
% with each other
through shared variables, message passing, or other means. In such
programs, the set of states reachable by one thread depends both on the
behaviors of the other threads,
% (i.e., what data they communicate to the first thread),
and on the order in which the threads are interleaved to create a global
execution. Since the thread interleaving is unknown to the program
designer, analysis techniques for asynchronous programs typically assume
the worst case, i.e., that threads can interleave arbitrarily; we refer to
this assumption as \emph{full scheduling nondeterminism}. In order to prove
safety properties of such programs, we must therefore ultimately
investigate all possible interleavings.

Proposed about a decade ago, \emph{delay-bounded deterministic
  scheduling}~\cite{DBLP:conf/popl/EmmiQR11} is an effective technique to
curb the large cost associated with exploring arbitrary thread
interleavings. The idea is that permitting a limited number of scheduling
\emph{delays}---skipping a thread when it is normally scheduled to
execute--- in an otherwise deterministic scheduler approximates a fully
nondeterministic scheduler from below.
% We say the scheduler delays a thread if it skips this thread at the time
% it would normally be scheduled to execute.
Delaying gives rise to a new thread interleaving, potentially
reaching states unreachable to the deterministic scheduler. In the limit,
i.e., with unbounded delays, the delaying and the fully nondeterministic
scheduler permit the same set of executions and, thus, reach the same
states.

Prior work has demonstrated that delay-bounded scheduling can ``discover
concurrency bugs efficiently''~\cite{DBLP:conf/popl/EmmiQR11}, in the sense
that such errors are often detected for a small number of permitted
delays. The key is that few delays means to explore only few
interleavings. Thus, under moderate delay bounds, the reachable state space
can often be explored exhaustively, resulting---if no errors are found---in
a delay-bounded verification result.
%% for a ``canonically characterized''~\cite{DBLP:conf/popl/EmmiQR11} fragment
%% of the set of reachable states.

We build on the empirical insight of efficient delay-\emph{bounded}
bug detection (testing) or verification, and make the following
contributions.

\emphasize{1.\ Delay-bounded scheduling without delay.} If no bug is found
while exhaustively exploring the given program for a given delay budget, we
``feel good'' but are left with an uncertainty as to whether the program is
indeed bug-free. We present a technique to remove this uncertainty, as
follows. We prove that the set $\reached[d]$ of states reached under a
delay bound $d$ equals the set $\reached$ of reachable states under
\emph{arbitrary} thread interleavings if two conditions are
met:
\begin{itemize}

\item increasing the delay bound by a number roughly equal to the number of
  executing threads produces no additional reachable states, and

\item set $\reached[d]$ is closed under a certain set of critical program
  actions.

\end{itemize}
In some cases, the set of ``critical program actions'' may be definable
statically at the language level; in others, this must be determined per
individual action. To increase the chance that the above two conditions
eventually hold, we typically work with conservative abstractions of
$\reached$; the (precisely computed) abstract reachability set
$\absreached$ is then used to decide whether the program is safe.

\emphasize{2.\ Efficient delay-unbounded analysis.} We translate the above
foundational result into an efficient delay-unbounded analysis
algorithm. It starts with a deterministic Round-Robin scheduler,
parameterized by the number of rounds $r$ it runs and of delays $d$ it
permits, and increases $r$ and $d$ in a delicate schedule \emph{weak-until}
the two conditions above hold\thomasmargin{this is not quite true, but
  close enough} (it is not guaranteed that they ever will). The key for
efficiency is that the reachability sets under increasing $r$ and $d$ are
\emph{monotone}. We therefore can determine reachability under parameters
$r' \atl r$ and $d' \atl d$ starting from a \emph{frontier} of the states
reached under bounds $r$ and~$d$. We present this algorithm and prove it
correct. We also prove its termination (either finding a bug or proving
correctness), under certain conditions.

\emphasize{3.\ Delay-unbounded analysis for general infinite-state
  systems.} We demonstrate the power of our technique on programs with
un\-bound\-ed-domain variables and unbounded thread counts. The existence
of in\-te\-ger-like variables suggests the use of a form of predicate
abstraction. Prior work has shown that predicate abstraction for
unbounded-thread concurrent programs leads to complex abstract program
semantics \cite{DKKW11,KKW17}, going beyond even the rich class of
well-quasiordered systems~\cite{A10}. Our delay-unbounded analysis
technique does not require an abstract program. Instead, we add to the idea
of reachability analysis under increasing $r$ and $d$ a third dimension
$n$, representing increasing thread counts, enjoying a similar convergence
property. Circumventing the static construction of the abstract program
simplifies the verification process dramatically.

In summary, this paper presents a technique to lift the bound used in
delay-bounded scheduling, while (empirically) avoiding the combinatorial
explosion of arbitrary thread interleavings. Our technique can therefore
find bugs as well as prove programs bug-free. We demonstrate its efficiency
using concurrent pushdown system benchmarks, as well as known-to-be-hard
infinite-state protocols such as the Ticket Lock~\cite{A91}. We offer a
detailed analysis of internal performance aspects of our algorithm, as well
as a comparison with several alternative techniques. We attribute the
superiority of our method to the retained parsimony of limited-delay
deterministic-schedule exploration.

\Ifthenelse{\arxiv}{Proofs omitted from the main text can be found in the
  Appendix, along with other supplementary information.}{A full version of
  this paper, with proofs omitted here and other supplementary information,
  can be found in an accompanying Technical
  Report~\cite{JW21a}.}\thomasmargin{\version}

\draftnewpage

\section{Delay-Bounded Scheduling}
\label{section: Delay-Bounded Scheduling}

\subsection{Basic Computational Model}
\label{section: Basic Computational Model}

For the purposes of introducing the idea behind delay-bounded scheduling,
we define a deliberately broad asynchronous program model. Consider a
multi-threaded program $\prg$ consisting of $n$ threads. We fix this number
throughout the paper up to and including \sectionref{The Fixed-Thread
  Case}, after which we consider parameterized scenarios. Each thread runs
its own procedure and communicates with others via shared program
variables.
% defined at the global level.
A~``procedure'' is a collection of \emph{actions} (such as those defined by
program statements). We define a shared-states set $G$ and, for each
thread, a local-states set $L_i$ ($0 \atm i < n$). A~global program state
is therefore an element of $G \times \Product_{i=0}^{n-1} L_i$. In
addition, a finite number of states are designated as
\emph{initial}. (Finiteness is required in \sectionref{Efficient
  Delay-Unbounded Analysis} for a termination argument [\lemmaref{DUBA
    termination}].)

The execution model we assume in this paper is
asynchronous. A~\emph{step} is a pair $(s,s')$ of states such
that there exists a thread $i$ ($0 \atm i < n$) such that $s$ and $s'$
agree on the local states of all threads $j \not= i$; the local state of
thread $i$ may have changed, as well as the shared state. We say thread $i$
\emph{executes} during the step, by executing some action of its
procedure.\footnote{If only the shared state changes, it is possible that
  the identity of the executing thread is not unique. This small ambiguity
  is inconsequential for this~paper.} The execution semantics within the
procedure is left to the thread (e.g., there may be multiple enabled
actions in a state, an action may itself be nondeterministic,
etc.). Without loss of generality for safety properties,
%, however,
we assume that the transition relation induced by each thread's possible
actions be total. That is, instead of an action $x$ being disabled for a
thread in state $s$, we stipulate that firing $x$ from $s$ results in $s$.

A \emph{path} is a sequence $p = (\range[]{s_0}{s_l})$ of states such that,
for $0 \atm i < l$, $(s_i,s_{i+1})$ is a step. This path has length $l$ (=
number of steps taken). A state~$s$ is \emph{reachable} if there exists a
path from some initial state to $s$. We denote by $\reached$ the (possibly
infinite) set of states reachable in $\prg$. Note that these definitions
permit arbitrary asynchronous thread interleavings.

\subsection{Free and Round-Robin Scheduling}

We formalize the notion of a scheduling policy indirectly, by
parameterizing the concept of reachability by the chosen scheduler. A state
$s$ is \emph{reachable under free scheduling} if there exists a path
$p=(\range[]{s_0}{s_l})$ from some initial state $s_0$ to $s_l = s$. A free
scheduler is simulated in state space explorers using full nondeterminism.
State $s$ is \emph{reachable under $n$-thread Round-Robin scheduling with
  round bound $r$}
% (``\emph{reachable under $\RR[r,n]$ scheduling}'' for short)
if there exists a path $p=(\range[]{s_0}{s_l})$ from some initial state
$s_0$ to $s_l = s$ such that
\begin{enumerate}

\item $\ceils{l/n} \atm r$, and

\item for $0 \atm i < l$, thread $i \pmod n$ executes during step $(s_i,s_{i+1})$.

\end{enumerate}

\subsection{Delay-Bounded Round-Robin Scheduling}

We approximate the set of states reachable under free
scheduling from below, using a relaxed Round-Robin scheduler. The scheduler
introduced so far is, however, deterministic and thus \emph{vastly}
underapproximates the free scheduler, even for unbounded $r$. The solution
proposed in earlier work is to introduce a limited number $d$ of scheduling
\emph{delays}~\cite{DBLP:conf/popl/EmmiQR11}. A delayed thread is skipped
in the current round and must wait until the next round.
\begin{DEF}
  \label{definition: reachable under RR with r and d}
  State $s$ is \emphdef{reachable under Round-Robin scheduling with round
    bound $r$ and delay bound $d$} (``\emph{reachable under $\RR[r,d]$
    scheduling}'' for short) if there exists a path
  $p=(\range[]{s_0}{s_l})$ from some initial state $s_0$ to $s_l = s$ and a
  function $\func f {\range 0 {l-1}} {\range 0 {n-1}}$, called \emph{scheduling function}, such that
  \begin{enumerate}

  \item \label{item: d_p}
    for $d_p := f(0) + \Sum_{i=1}^{l-1} \big( (f(i) - f(i-1) - 1) \bmod n \big)$, we have $d_p \atm d$,

  \item \label{item: r bound}
    $\ceils{\frac{l + d_p} n} \atm r$ ($d_p$ as defined in \itemref[]{d_p}.), and

  \item \label{item: scheduling constraint}
    for $0 \atm i < l$, thread $f(i)$ executes during step $(s_i,s_{i+1})$.

  \end{enumerate}
\end{DEF}
Variable $d_p$ from
\itemref[]{d_p}.~quantifies the total delay, compared to a perfect
Round-Robin scheduler, that the scheduling along path $p$ has
accumulated. Consider the case of $n=4$ threads T0,\ldots,T3. Then the
scheduling sequence $(\range[]{f(0)}{f(11)})$ below on the left, of $l=12$
steps and involving 13 states, follows a perfect Round-Robin schedule of
$r=3$ rounds (separated by~\verb:|:):\format
\renewcommand{\localCommand}[1]{\hspace*{#1pt}{\color{red}\huge\texttimes}}
\[
\begin{tabular}{rcl}
  \verb:0 1 2 3 | 0 1 2 3 | 0 1 2 3: & \hspace*{20pt} & \verb:0 1 2 3 | 0 1 2 3 | 0 1 2 3: \\[-12.5pt]
                                     &                & \localCommand{16.5}\localCommand{48}\localCommand{6}
\end{tabular}
\]
The sequence on the right of $l=9$ steps follows a Round-Robin scheduling
of $r=3$ rounds and a total of $d_p = 3$ delays: one after the second step
(T2 is delayed: $3-1-1 \bmod 4 = 1$), another two delays after the sixth
step (T3 and T0 are delayed: $1-2-1 \bmod 4 = 2$). The final state of this
path is reachable under $\RR[3,3]$ scheduling. Note that delays effectively
shorten rounds\Ifspace{\ (the scheduling position is ``wasted'')}.

We denote by $\reached[r,d]$ the set of states reachable in $\prg$ under
$\RR[r,d]$ scheduling. (Note that this set is finite, for any program
$\prg$.) It is easy to see that, given sufficiently large $r$ and $d$,
\emph{any} schedule can be realized under $\RR[r,d]$ scheduling:
\newcounterset{DUBAisFREE}{\theASS}
\begin{THE}
  \label{theorem: DUBA = FREE}
  State $s$ is reachable under free scheduling iff there exist $r,d$ such
  that $s$ is reachable under $\RR[r,d]$ scheduling: $\reached =
  \Union_{r,d \in \NN} \reached[r,d]$.
\end{THE}
State-space exploration under free scheduling can therefore be reduced to
enumerating the two-dimensional parameter space $(r,d)$ and computing
states reachable under $\RR[r,d]$ scheduling. This can be used to turn a
Round Robin-based state explorer into a semi-algorithm, dubbed
\emph{delay-bounded tester} in~\cite{DBLP:conf/popl/EmmiQR11}.

An important property of the round and delay bounds is that increasing them
can only increase the reachability sets:
\begin{PRO}[Monotonicity in $r$ \& $d$]
  \label{property: monotonicity in r and d}
  For any round and delay bounds $r$ and $d$:
  \begin{equation}
    \label{equation: monotonicity in r and d}
    \reached[r,d] \subseteq \reached[r+1,d] \wbox{,} \reached[r,d] \subseteq \reached[r,d+1] \ .
  \end{equation}
\end{PRO}
This follows from the $\ldots \atm r$ and $\ldots \atm d$ constraints in
\definitionref{reachable under RR with r and d}. The property relies on $r$
and $d$ being external to the program, not accessible inside~it.
% (or in any way alias to variables in $\prg$).
%% Note that, in contrast, $\reached[r,d,n] \subseteq \reached[r,d,n+1]$ is
%% \emphasize{not} valid in general; see \sectionref{The Unbounded-Thread
%%   Case}.
Under this provision, monotonicity in any kind of resource bound is a fairly
natural \emph{yet not always guaranteed} property; we give a counterexample
in \sectionref{The Unbounded-Thread Case}.

\Ifspace{It is interesting to ponder the different contributions of $r$ and
  $d$ to the set of arbitrary thread schedules. In general, both parameters
  need to be allowed to grow without bound to simulate a free
  scheduler. For example, bounding $r$ by $r_0$ causes states with distance
  greater than $r_0 \cdot n$ to be unreachable, no matter what $d$. This
  permits only finitely many reachable states and is thus incomplete for
  infinite-state systems.

\thomas{Q: I wonder about the following: is there a $d_0$ such that
  \[
    \lim_{r \rightarrow \infty} \RR[r,d_0] = \mathit{FREE} \ ?
  \]
  Probably not, but I don't have an easy argument.}

\andrew{We know that as long as the rounds are large enough, increasing the
  delay is the important parameter. \thomas{can we explain what ``important
    parameter'' means?}}\andrew{My use of important here means that it contributes a lot more schedules when increased; i.e. i trust a plateau with a lot of delays more than a plateau with a lot of rounds.}
}

\Ifspace{Round-Robin scheduling can additionally be parameterized by the
  number of transitions a thread can take during its turn. This parameter
  is often called the \emph{slice size}. In this paper, we assume a fixed
  slice size of 1. We can afford this liberty since larger slice sizes can
  be simulated using larger round and delay bounds. Consider a thread $T$
  that ``wishes'' to take an extra step before being preempted by the
  scheduler. We can accomplish this by delaying all other threads and
  waiting for the next round, effectively retaining control with~$T$. This
  may require an increased number of delays (by $n-1$) and rounds (by~1).}

\andrew{So, in the limit where there are infinite delays/rounds, the slice
  size is not actually a limit; any positive constant will work. This means
  that the user can choose a slice size empirically based on their program
  that is big enough to let every machine do everything interesting it can
  in one turn.}

\section{Abstract Closure for Delay-Bounded Analysis}
\label{section: Abstract Closure for Delay-Bounded Analysis}

The goal of this paper is a technique to prove safety properties of
asynchronous programs under arbitrary thread schedules. \theoremref{DUBA =
  FREE} affords us the possibility to reduce the exploration of such
arbitrary schedules to certain bounded Round-Robin schedules, but we still
need to deal with those bounds. In this section we present a closure
property for bounded Round-Robin explorations.

%% \andrewmargin{this could be a little more succinct as well if we have an
%%   overview that lays out what is in each section earlier, rather than
%%   restating future section goals}

\subsection{Respectful Actions}
\label{section: Respectful Actions}

Let $\concStates$ be the set of global program states of $\prg$, and let
$\func{\alpha}{\concStates}{\abstrStates}$ be an \emph{abstraction
  function}, i.e., a function that maps program states to elements of 
some abstract domain $\abstrStates$. Function $\alpha$ typically hides
certain parts of the information contained in a state, but the exact
definition is immaterial for this subsection.

A key ingredient of the technique proposed in this paper is to identify
actions of the program executed by a thread with the property that the
abstract successor of an abstract state under such an action does not
depend on concrete-state information hidden by the abstraction.
\begin{DEF}
  \label{definition: respectful action}
  Let $x$ be a program action, and let the relation $\step s i x {s'}$
  denote that $s \rightarrow s'$ is a step during which thread $i$ executes
  $x$. Action $x$ \emphdef{respects} $\alpha$ if, for all states
  $s_1,s_2,s_1',s_2' \in \concStates$ and all $i: 0 \atm i < n$:
  \begin{equation}
    \label{equation: respectful action}
    \alpha(s_1) = \alpha(s_2)                               \ \land \
    s_1 \stackrel{\mboxscript{$i,x$}}{\longrightarrow} s_1' \ \land \
    s_2 \stackrel{\mboxscript{$i,x$}}{\longrightarrow} s_2' \wbox{$\limplies$}
    \alpha(s_1') = \alpha(s_2') \ . 
  \end{equation}
%  An action that does not respect $\alpha$ \emphdef{disrespects} $\alpha$.
\end{DEF}
%% Also this can be an IFF, does it matter?\thomas{\ no, this cannot be an
%%   IFF, think about it. With our visible-state projection~$\alpha$, what
%%   if $s_1=(0|\emptystack)$, $s_2=(1|\emptystack)$ (so $\alpha(s_1) \not=
%%   \alpha(s_2)$), and $x$ is an action that changes the shared state to 1
%%   (which is very respectful). Then $\alpha(s_1') = \alpha(s_2')$}
%%
%% \thomasmargin{Alternative terminology: ``well-defined'' is not the right
%%   notion since we do not apply a function to an equivalence class
%%   here. ``$\alpha$ invariant under $x$-images'' is too strong: it means
%%   $\alpha(Im_x(s)) = \alpha(s)$.}
Intuitively, ``$x$ respects $\alpha$'' means that successors under action
$x$ of $\alpha$-equiv\-alent states all have the same unique
abstraction.\Ifspace{\ Equivalently, it means that the abstract successor
  relation under action $x$ is a function.} Note the special case $s_1 =
s_2$, $s_1' \not= s_2'$: for nondeterministic actions $x$ to
respect~$\alpha$, multiple successors $s_1',s_2'$ of the same concrete
state $s_1 = s_2$ under $x$ also must have the same abstraction.
\begin{EXA}
  \label{example: DUBA in CPDS}
  Consider $n$-thread \emph{concurrent pushdown systems} (CPDS), an
  instance of the asynchronous computational model presented in
  \sectionref{Basic Computational Model}. We have a finite set of shared
  states readable and writeable by each thread. Each thread also has a
  finite-alphabet stack, which it can operate on by (i) overwriting the
  top-of-the-stack element, (ii) pushing an element onto the stack, or
  (iii) popping an element off the top of the non-empty stack. The classic
  pointwise top-of-the-stack abstraction function is defined by
  \begin{equation}
    \label{equation: alpha}
    \alpha(g,\range[]{w_0}{w_{n-1}}) = (g,\range[]{\sigma_0}{\sigma_{n-1}}) \ ,
  \end{equation}
  where $g$ is the shared state (unchanged by $\alpha$), $w_i$ is the
  contents of the stack of thread $i$, and $\sigma_i$ is the top of $w_i$
  if $w_i$ is non-empty, and empty otherwise~\cite{LW18}. Note that the
  domain into which $\alpha$ maps is a finite set.

  Push and overwrite actions respect $\alpha$, while pop actions disrespect
  it: consider the case $n=1$ and $s_1 = (g,w_0) = (0,10)$ and $s_2 =
  (0,11)$, with stack contents~$10$ and~$11$, resp.
% (left to right =
  (left = top). While $\alpha(s_1) = \alpha(s_2) = (0,1)$, the (unique)
  successor states of $s_1$ and $s_2$ after a \emph{pop} are not
  $\alpha$-equivalent: the elements $0$ and $1$ emerge as the new
  top-of-the-stack symbols, respectively, which $\alpha$ can distinguish.
\end{EXA}

The notion of respectful actions gives rise to a condition on sets of
abstract states that we will later use for convergence proofs:
\renewcommand{\localCommand}{\emphasize a}
\begin{DEF}
  \label{definition: closed}
  An abstract-state set $A$ is \emphdef{closed under actions
    disrespecting~$\alpha$} if,
% $A$ is closed under computing successors under disrespectful actions,
% i.e.\ if,
  for every $\localCommand \in A$ and every successor $\localCommand'$ of
  $\localCommand$ under a disrespectful action, $\localCommand' \in
  A$.
\end{DEF}
For maximum precision: $a'$ is said to be a successor of $a$ under a
disrespectful action if there exist concrete states $s$ and $s'$, a~thread
id $i$ and an action $x$ such that $\alpha(s) = a$, $\alpha(s') = a'$,
$x$~disrespects $\alpha$, and~$\step s i x {s'}$. If abstraction $\alpha$
is clear from the context, we may just say ``closed under disrespectful
actions''.

\subsection{From Delay-Bounded to Delay-Unbounded Analysis}
\label{section: From Delay-Bounded to Delay-Unbounded Analysis}

We now present our idea to turn a round- and delay-bounded tester into a
(partial) verifier, namely by exploring the given asynchronous program for
a number of round and delay bounds until we have ``seen enough''. Recall
the notations $\reached$ and $\reached[r,d]$ defined in
\sectionref{Delay-Bounded Scheduling}. We also use $\absreached$ and
$\absreached[r,d]$ short for $\alpha(\reached)$ and
$\alpha(\reached[r,d])$, i.e.\ the respective abstract reachability
sets. (Note that $\absreached$ is \emph{not} an abstract fixed
point---instead, it is the result of applying $\alpha$ to the concrete
reachability set $\reached$; see discussion in \sectionref{Discussion of
  Related Work}.)
\begin{THE}
  \label{theorem: DUBA}
  For any $r,d \in \NN$, if $\absreached[r,d] = \absreached[r+1,d+n-1]$ and
  $\absreached[r,d]$ is closed under actions disrespecting $\alpha$, then
  $\absreached[r,d] = \absreached$.
\end{THE}
The theorem states: if the set of \emph{abstract} states reachable under
$\RR[r,d]$ scheduling does not change after increasing the round bound by 1
and the delay bound by $n-1$, and it is closed under disrespectful actions,
then $\absreached[r,d]$ is in fact the \emph{exact} set $\absreached$ of
abstract states reachable under a \emph{free} scheduler: no approximation,
no rounds, no delays, no Round-Robin.

\Paragraph

\Proof\ of \theoremref{DUBA}: we have to show that $\absreached[r,d]$ is closed under the
abstract image function $\absIm$ induced by $\alpha$, defined as
\[
  \absIm(a) \ = \ \{a': \ \exists s,s' : \ \alpha(s) = a, \ \alpha(s') =
  a', \ s \rightarrow s' \} \ .
\]
That is, we wish to show $\absIm(\absreached[r,d]) \subseteq
\absreached[r,d]$, which proves that no more abstract states are
reachable. Consider $a \in \absreached[r,d]$ and $a' \in \absIm(a)$,
i.e.\ we have states $s,s'$ such that $\alpha(s) = a$, $\alpha(s') = a'$,
and $\step s i x {s'}$ for some thread~$i$ and some action~$x$. The goal is
to show that $a' \in \absreached[r,d]$.

To this end, we distinguish flavors of $x$. If $x$ disrespects $\alpha$,
then $a' \in \absreached[r,d]$, since the set is closed under disrespectful
actions.

So $x$ respects $\alpha$. Since $a \in \absreached[r,d]$, there exists a
state $s_0 \in \reached[r,d]$
%% \andrew{\ just use the $s$ and $s'$ from above}\thomas{\ actually, we
%%   cannot use either. We only know $s \rightarrow s'$ and
%%   $\alpha(s)=a$. We do not know that $s \in \reached[r,d]$, which we
%%   need here}
with $\alpha(s_0) = a$. Suppose for a moment that thread $i$ is scheduled
to run in state $s_0$. Then it can execute action $x$; any successor state
$s_0'$ satisfies $s_0' \in \reached[r,d]$, and:
\begin{equation*}
  a' \opNote{(def $a'$)} = \alpha(s') \opNote{($x$ resp.~$\alpha$)} = \alpha(s_0') \opNote{(def $s_0'$)}{\in} \alpha(\reached[r,d]) = \absreached[r,d] \ .
\end{equation*}
\emphasize{But what if} the thread scheduled to run in state $s_0$ under
$\RR[r,d]$ scheduling, call it $j$, \emphasize{is not thread $i$?} Then we
\emph{delay} any threads that are scheduled before thread $i$'s next turn;
if $i < j$, this ``wraps around'', and we need to advance to the next
round. The program state has not changed---we are still in $s_0$. Let
$s_0'$ be the successor state obtained when thread $i$ now executes action
$x$, and $\lambda(i,j) = 1$ if $i<j$, 0 otherwise. Then we have $s_0' \in
\reached[r+\lambda(i,j), d+(j-i) \bmod n]$, and:
\begin{equation*}
  \begin{array}{rcl}
    a' \opNote{(def $a'$)} = \alpha(s') \opNote{($x$ resp.~$\alpha$)} = \alpha(s_0') & \opNote{(def $s_0',\alpha$)}{\in}    & \absreached[r+\lambda(i,j), d+(j-i) \bmod n] \\
                                                                                     & \opNote{(monot.~$r$,$d$)}{\subseteq} & \absreached[r+1,d+n-1] \opNote{(\theoremref{DUBA})}= \absreached[r,d] \ .
  \end{array}
\end{equation*}
This concludes the proof of \theoremref{DUBA}.\eop

%% The proof idea is as follows. Due to the closure property, we only need to
%% consider respectful actions $x$ that lead to abstract successors $a'$ of $a
%% \in \absreached[r,d]$, say via thread~$i$. If~$i$ can execute $x$ in the
%% concrete state $s_0$ that witnesses the reachability of~$a$, then---due to
%% respectfulness---the abstraction of the successor equals $a'$, from which
%% $a' \in \absreached[r,d]$ follows. But what if $i$ is not scheduled to
%% execute in state $s_0$? Then we delay the next $n-1$ threads, which may
%% take us into the next RR-round; we then have $a' \in
%% \absreached[r+1,d+(n-1)] = \absreached[r,d]$.
\begin{EXA}
  \label{example: long delays}
  Consider a simple 3-thread system with a shared-states set $G =
  \{0,1,2\}$. The local state of each thread is immaterial; function
  $\alpha$ just returns the shared state: $\alpha(g,l_0,l_1,l_2) = g$. The
  threads' procedures consist of the following actions, which update only
  the shared state:
  \begin{quote}
    Thread T0: $0 \rightarrow 1$ \hfill Thread T1: $0 \rightarrow 1$  \hfill Thread T2: $0 \rightarrow 2$ \ .
  \end{quote}
\tableref{long delays} shows the set of reachable states for different
round and delay bounds.  For example, with one round and zero delays, the
only feasible action is T0's. The reachable states are $0$ (initial) and
$1$ (found by T0). The table shows a path to a pair $(r,d)$ that meets the
conditions of \theoremref{DUBA}. From $(r,d)=(1,0)$ we increment $r$ to
find a plateau in $r$ of length $1$. We then increase $d$ to try to find a
plateau in $d$ of length $n-1=2$. This example shows that a delay plateau
of length $1$ is not enough, as $3$ is only reachable with at least $2$
delays. At $(2,2)$ we find a new state ($3$), so we restart the search for
plateaus in $r$ and $d$. At $(3,4)$, the plateau conditions for
\theoremref{DUBA} are met. There are no disrespectful transitions, so by
\theoremref{DUBA}, we know that $\absreached[3,4]=\absreached$.
\end{EXA}
\begin{table}[htbp]
  \caption{Reachable states in \exampleref{long delays} under various round
    and delay bounds.
% Arrows show a path through this space.
    The boxed set passes the convergence test suggested by
    \theoremref{DUBA}}
  \label{table: long delays}
  \begin{center}
    \bgroup
    \def\arraystretch{1.5}%
    \begin{tabularx}{234pt}{|r|ccccc|}
      \hline
            & \hspace*{3mm}$d=0$\hspace*{3mm} & \hspace*{3mm}$d=1$\hspace*{3mm} & $\hspace*{3mm}d=2\hspace*{3mm}$ & $\hspace*{3mm}d=3\hspace*{3mm}$ & $\hspace*{3mm}d=4$\hspace*{3mm} \\ \hline
%     $r=0$ & \tikz[baseline]{\node (p1)  {\{0\}}} & \{0\} & \{0\} & \{0\} & \{0\} \\[-4pt]
      $r=1$ & \tikz[baseline]{\node (p1)  {\{0,1\}}} & \{0,1\} & \{0,1,2\} & \{0,1,2\} & \{0,1,2\} \\
      $r=2$ & \tikz[baseline]{\node (p2)  {\{0,1\}}} & \tikz[baseline]{\node (p3)  {\{0,1\}}} & \tikz[baseline]{\node (p4)  {\{0,1,2\}}} & \tikz[baseline]{\node (p5)  {\{0,1,2\}}} & \tikz[baseline]{\node (p6)  {\{0,1,2\}}} \\
      $r=3$ & \tikz[baseline]{\node (p7)  {\{0,1\}}} & \tikz[baseline]{\node (p8)  {\{0,1\}}} & \tikz[baseline]{\node (p9)  {\{0,1,2\}}} & \tikz[baseline]{\node (p10) {\{0,1,2\}}} & \tikz[baseline]{\node[draw] (p21) (p11) {\{0,1,2\}}} \\ \hline
    \end{tabularx}
    \egroup
    \begin{tikzpicture}[overlay]
      \path[thick,->] (p1) edge (p2) ;
      \path[thick,->] (p2) edge (p3) ;
      \path[thick,->] (p3) edge (p4) ;
      \path[thick,->] (p4) edge (p9) ;
      \path[thick,->] (p9) edge (p10) ;
      \path[thick,->] (p10) edge (p11) ;
    \end{tikzpicture}
  \end{center}
\end{table}

\draftnewpage

\section{Efficient Delay-Unbounded Analysis}
\label{section: Efficient Delay-Unbounded Analysis}

Turning \theoremref{DUBA} into a reachability algorithm requires efficient
computation of the sets $\reached[r,d]$. This section presents an approach
to achieve this, by expanding only \emph{frontier} states when either the
round or the delay parameter is increased.

To this end, let $\prop$ be a state property (such as an assertion) that
respects~$\alpha$, in the sense that, for any states $s_1$, $s_2$, if
$\alpha(s_1) = \alpha(s_2)$, then $s_1 \models \prop$ iff $s_2 \models
\prop$. From now on, we further assume the domain $\abstrStates$ of
abstraction function $\alpha$ to be finite, which will ensure termination
of our algorithm (see \lemmaref{DUBA termination} later).

Our verification scheme for $\prop$ is shown in \algorithmref{DUBA}, which
uses \algorithmref{FinishRounds} as a subroutine. In the rest of this
paper, we also refer to \algorithmref{DUBA} as \emph{Delay-(and
  round-)\format\linebreak UnBounded Analysis}, \duba\ for
short.\andrewmargin{we need to exclude states that had max rounds in line
  17, it is an annoying edge case}
\begin{algorithm}[htbp]
  \begin{algorithmic}[1]
    \REQUIRE{$n$-thread asynchronous program, property $\prop$}
    \ENSURE{``safe'', ``violation of $\prop$'', or ``unknown''}
    \STMT \code{$\Reached$ := \textnormal{(finite) set of initial states}} \COMMENT{$\Reached$: states reached so far} \label{line: initial states}
    \STMT \code {$r$ := 0; $d$ := 0}                                            \label{line: initialize r,d}
    \REPEAT                                                                     \label{line: round loop begin}
      \STMT \code{$\Frontier$ := $\{s \in \Reached: s.\roundsTaken = r\}$}
      \STMT \code{$r$++}
      \FOR {$s \in \Frontier$}                                                  \label{line: for loop round begin}
        \STMT \code{$\Reached$ := $\Reached \union \FinishRounds(s,r+1,\prop)$} \label{line: for loop round body}
      \ENDFOR
      \STMT \code{$r$++}
    \UNTIL round plateau of length 1                                            \label{line: round loop end}
    \REPEAT                                                                     \label{line: delay loop begin}
      \STMT \code{$\Frontier$ := $\{s \in \Reached: s.\delaysTaken = d\}$}      \label{line: delay frontier}
      \STMT \code{$d$++}                                                        \label{line: delay increment}
      \FOR {$s \in \Frontier$}                                                  \label{line: for loop delay begin}
        \STMT \code{$s'$ := $s$} \COMMENT{copy of state $s$}
	\STMT \code{$s'.\delaysTaken$++}                                        \label{line: delay taken}
	\STMT \code{$s'.\finder$ := ($s'.\finder$ + 1) mod n}
	\IF {\code{$s'.\finder$ mod n = 0}}
	  \STMT \code{$s'.\roundsTaken$++}
	\ENDIF
	\STMT \code{$\Reached$ := $\Reached \union \FinishRounds(s',r,\prop)$}  \label{line: finishrounds delay}
      \ENDFOR
      \IF{new abstract state found during \plfor\ loop in \lineref{for loop delay begin}}  \label{line: check delayed state new}
        \STMT \plgoto\ \lineref[]{round loop begin} \COMMENT{abort second \plrepeat\ loop; go back to first} \label{line: start over}
      \ENDIF
    \UNTIL delay plateau of length $n-1$                                        \label{line: delay loop end}
    \IF {$\alpha(\Reached)$ is closed under disrespectful actions}              \label{line: convergence test}
      \STMT \plreturn\ ``safe''                                                 \label{line: safe}
    \ELSE
      \STMT \plreturn\ ``unknown''
    \ENDIF
  \end{algorithmic}
  \caption{Verifying property $\prop$ against all reachable states of program $\prg$}
  \label{algorithm: DUBA}
\end{algorithm}

\begin{algorithm}[htbp]
  \begin{algorithmic}[1]
    \REQUIRE{$s$: state, $r$: round bound, $\prop$: state property}
    \ENSURE{states reachable from $s$ up to round bound $r$, without delaying}
    \STMT \code{Set<State> $\Unexplored$ := $\{s\}$, $\Reached$ := $\{\}$}
    \WHILE {\code{$\Unexplored$ != $\{\}$}}
      \STMT select and remove some state $u$ from $\Unexplored$
      \IF {$u$ violates $\prop$}
        \STMT throw ``violation of $\prop$ (witnessed by reaching state $u$)'' \label{line: violation}
      \ENDIF
      \STMT \code{$\Reached$ := $\Reached \union \{u\}$}
      \IF[if $u$ schedulable]{\code{$u.\finder < n-1$ or $u.\roundsTaken < r$}}
        \STMT \code{$\Unexplored$ := $\Unexplored \union \left(\Image(u) \setminus \Reached\right)$} \label{line: image computation}
    \ENDIF
    \ENDWHILE
    \STMT \plreturn\ $\Reached$
  \end{algorithmic}
  \caption{$\FinishRounds(s,r,\prop)$}
  \label{algorithm: FinishRounds}
\end{algorithm}

\draftnewpage

The main data structure used in the algorithms is that of a \code{State},
which stores both program variables and scheduling information, in the
attributes $\finder$, $\roundsTaken$, and $\delaysTaken$. For a state $s$,
variables $s.\roundsTaken$ and $s.\delaysTaken$ represent the number of
times the scheduler started a round and delayed a thread, resp., to get to
$s$. Variable $s.\finder$ contains the index of the thread whose action
produced $s$. This is enough information to continue the execution from $s$
later, starting with the thread after $\finder$. For the initial states,
$\roundsTaken$ and $\delaysTaken$ are zero, and $\finder$ is $n-1$ (the
latter so that expanding the initial states starts with thread $(n-1)+1
\bmod n = 0$). For set membership testing, two states are considered equal
when they agree on their finders and on program variables. The
$\roundsTaken$ and $\delaysTaken$ variables are for scheduling purposes
only and ignored when checking for equality.

As mentioned in \propertyref{monotonicity in r and d}, the sequence of
reachability sets is monotone with respect to both rounds and delays, for
any program. This entails two useful properties for
\algorithmref{DUBA}. First, we can increase the bounds in any order and at
individual rates.\Ifspace{\ As long as we are going ``out'' (increasing
  either parameter by at least one), we are (weakly) increasing the number
  of schedules explored.} Second, it suffices to expand states at the
frontier of the exploration, without missing new schedules. When adding a
new delay, we only need to delay those states that were (first) found in
schedules using the maximum delays. When adding a round, we only need to
expand states that were (first) found in the last round of a schedule.

\algorithmref{DUBA} first advances the round parameter $r$ until a
round plateau has been reached (Lines~\lineref[]{round loop
  begin}--\lineref[]{round loop end}). It does so by running the
$\FinishRounds$ function on \emph{frontier states} $s$: those that were
reached in the final round $r$ of the previous round iteration.
%, i.e., in round $r-1$.
$\FinishRounds$ (\algorithmref{FinishRounds}) explores from the given state
$s$, Round-Robin style, up to the given round, without delaying any
thread. The actual expansion of a state happens in function $\Image$
(\lineref{image computation} of \algorithmref{FinishRounds}), which computes a state's successors and
initializes their scheduling variables: $\roundsTaken$ and $\delaysTaken$
are copied from $u$, the $\finder$ of the successor is the next thread ($+1
\bmod n$). If this wraps around, $\roundsTaken$ is incremented as well.

Back to the main \algorithmref{DUBA}: we have reached a round plateau of
length 1 if the entire $\plfor$ loop in \lineref{for loop round begin} sees
no new \emph{abstract} states (no new elements in $\alpha(\Reached)$). If
so, we are not ready yet to perform the convergence test (recall
\exampleref{long delays}). Instead, \algorithmref{DUBA} now similarly
advances the delay parameter $d$ (Lines~\lineref[]{delay loop
  begin}--\lineref[]{delay loop end}). For each frontier state
($\delaysTaken = d$), we delay the thread scheduled to execute from this
state (by incrementing (mod $n$) the $\finder$ variable), and record the
taken delay (\lineref{delay taken}). Then we again call the $\FinishRounds$
function and merge in the states found. Importantly, these merges preserve
states already in $\Reached$, meaning that the algorithm will keep states
found earlier in the exploration (with smaller $r,d$).

The loop beginning in \lineref{delay loop begin} repeats until a delay
plateau of length $n-1$ is encountered (as required by
\theoremref{DUBA}). This means that during $n-1$ consecutive
\plrepeat\ iterations, the \plfor\ loop in \lineref[]{for loop delay begin}
did not find any new abstract states. When the round and delay plateaus
have the required lengths (1 and $n-1$, resp.), we invoke the convergence
test (\lineref{convergence test}), which amounts to applying
\theoremref{DUBA}. If the test fails, \algorithmref{DUBA} returns
``unknown''.

Towards proving partial correctness of \algorithmref{DUBA}, we first show
that the states eventually collected in set $\Reached$ by the algorithm
correspond exactly to the round- and delay-bounded reachability sets
$\reached[r,d]$, and that---after the two main \plkeyword{repeat} loops---a
plateau of sufficient length has been generated. As a corollary, the
algorithm is partially correct, i.e. it returns correct answers if it
terminates.
\newcounterset{PlateauCondition}{\theASS}
\begin{LEM}
  \label{lemma: plateau condition}
  If \algorithmref{DUBA} reaches \lineref{convergence test}, the current
  values of $r$ and $d$ satisfy: \\
  (i)~$\Reached = \reached[r,d]$, and (ii) $\absreached[r-1,d-(n-1)] =
  \absreached[r,d]$.
\end{LEM}
\newcounterset{DUBAsoundness}{\theASS}
\begin{COR}
  \label{corollary: DUBA soundness}
  The answers ``safe'' and ``violation of $\prop$'' returned by
  \algorithmref{DUBA} are correct.
\end{COR}

The algorithm won't return either ``safe'' or ``violation of $\prop$'' in
one of two situations: when the convergence test fails in
\lineref{convergence test} (it gives up), and when it fails to ever reach
this line. The latter can be prevented using a finite-domain $\alpha$:
\newcounterset{DUBAtermination}{\theASS}
\begin{LEM}
  \label{lemma: DUBA termination}
  If the domain $\abstrStates$ of abstraction function $\alpha$ is finite,
  \algorithmref{DUBA} terminates on every input.
\end{LEM}
Since abstraction $\alpha$ approximates the information contained in a
state, a plateau may be \emph{intermediate}, e.g.\ $\absreached[1,0]
\subsetneq \absreached[1,1] = \absreached[2,2] \subsetneq
\absreached[2,3]$. Thus, stopping the exploration simply on account of
encountering a plateau---even of lengths $(1,n-1)$---is
unsound. Intermediate plateaus make our algorithm (unavoidably) incomplete:
if the test in \lineref{convergence test} fails, then there are
known-to-be-reachable abstract states with abstract successors whose
reachability cannot be decided at that moment. If we knew the plateau to be
intermediate, we could keep exploring the sets $\reached[r,d]$ for larger
values of $r$ and $d$ until the next plateau emerges, hoping that the
convergence test succeeds at that time. In general, however, we cannot
distinguish intermediate from final plateaus.
%% This seems to be very rarely successful in practice, so our algorithm
%% simply gives up in this case.

\draftnewpage

\section{\duba\ with Unbounded-Domain Variables}
\label{section: DUBA with Unbounded-Domain Variables}

In addition to unbounded control structures like stacks, which come up in
pushdown systems and were discussed in \exampleref{DUBA in CPDS}, infinite
state spaces in programs are often due to (nominally) unbounded-domain
program variables. This presents no problem for the computation of the
concrete reachability sets $\Reached$ in \algorithmref{DUBA}: for any round
and delay bounds $(r,d)$, the set of concrete reachable states $\RR[r,d]$
is finite and thus explicitly computable (no symbolic data structures are
needed).\footnote{Contrast this to a \emph{context-switch} bound, under
  which reachability sets can be infinite.} On the other hand, termination
of the same algorithm requires that it eventually reach a plateau in $r$
and $d$ of sufficient length. This is guaranteed by an abstraction function
$\alpha$ that maps concrete states into a finite abstract space. A finite
abstract domain is therefore highly desirable.

A generic abstraction that reduces an unbounded data domain to a finite one
is predicate abstraction
\Ifthenelse{\arxiv}{(\appendixref{Predicate-Abstraction Primer} offers a
  short primer)}{\cite[see~\cite{JW21a} for a short
    primer]{GS97,BMMR01}}\thomasmargin{\version}. The goal in this section
is to demonstrate how the simple scheme of delay-unbounded analysis can be
combined with predicate abstraction to verify unbounded-thread programs.

\draftnewpage

\subsection{The Fixed-Thread Case}
\label{section: The Fixed-Thread Case}

Consider program $\otherprg$ in \figureref{toy} on the left \cite[page 4:
  program~$\mathbb P''$]{DKKW11}. Intuitively, variable $m$ counts the
number of threads spawned to execute~$\otherprg$ concurrently. It is easy
to see that ``the assertion in [$\otherprg$] cannot be violated, no matter
how many threads execute [$\otherprg$], since no thread but the first will
manage to''~\cite{DKKW11} enter the $\true$ branch of the \plif\ statement
and reach the assertion.
\begin{figure}[htbp]
  \centerTwoOut{%
    \fbox{%
      \begin{minipage}[c]{.4\textwidth}
        \renewcommand{\algorithmicrequire}{}
        Program $\otherprg$ \\[-.5\baselineskip]
        \horiline \\[-\baselineskip]\format
        \begin{algorithmic}[1]
          \setcounter{ALC@line}{-1} % to start numbering with 0
          \REQUIRE{\plkeyword{shared int} \code{$m$ := 0}}
          \REQUIRE{\plkeyword{shared int} \code{$s$ := 0}}
          \REQUIRE{\plkeyword{local  int} \code{$l$ := 0}}
          \STMT \code{$m$++}                                 \label{line: toy: inc m}
          \IF {\code{$m$ = 1}}                               \label{line: toy: if branch}
            \STMT \code{$s$++, $l$++}                        \label{line: toy: double increment}
            \STMT \plassert\ \code{$s$ = $l$}                \label{line: toy: assertion}
            \STMT \plgoto\ \lineref[]{toy: double increment} \label{line: toy: goto}
          \ENDIF
        \end{algorithmic}
      \end{minipage}
    }}{
    \begin{minipage}[c]{.5\textwidth}
      \includegraphics[scale=0.7]{\iffalse Images/ \fi 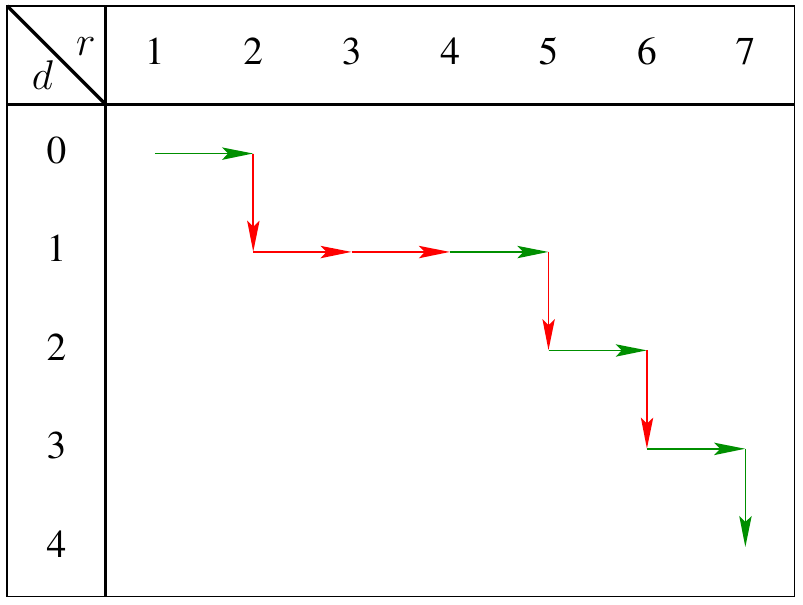}
    \end{minipage}
  }
  \caption{%
    \emphasize{Left:} program $\otherprg$;
    \emphasize{Right:} how \algorithmref{DUBA} operates on (an abstraction of) it}
  \label{figure: toy}
\end{figure}

Previous work has shown that even the 1-thread version of this program
cannot be proved correct using predicate abstraction \emphasize{unless} we
permit predicates that depend on both shared and local variables \cite[for
  the unprovability result]{DKKW11a}, which have been referred to as
\emph{mixed}~\cite{DKKW11}. An example is the predicate \mbox{$p :: (s =
  l)$}, which comes up in the assertion. The dependence of $p$ on both
shared and thread-local data causes standard solutions that track the truth
value of $p$ in a shared or local Boolean variable to be unsound. The
solution proposed in~\cite{DKKW11} is to
% track it in a local Boolean variable and
use \emph{broadcast} instructions to have the executing thread notify all
other threads whenever the truth value of $p$ changes. This solution comes
with two disadvantages: (i) the resulting Boolean broadcast programs are
more expensive to analyze than strictly asynchronous Boolean programs, and
(ii) the solution cannot be extended to the unbounded-thread case.

\Paragraph

Let us consider how we can verify this program using \algorithmref{DUBA}, 
for the fixed-thread case; we consider $n=2$ threads.
%, the smallest case relevant under concurrency.
We will have to use mixed predicates as in~\cite{DKKW11}, but since we
never execute the abstract Boolean program, there is no need for
constructing it. As a result, there is no need for broadcast instructions.

The program generates an unbounded number of reachable concrete states, but
we explore it only under round and delay bounds $r$ and $d$. As per
\algorithmref{DUBA}, we increase these bounds until we have reached
plateaus of lengths 1 and \mbox{$n-1=1$}, resp. Plateaus are determined
over the abstract-state set, so we need a function~$\alpha$.

\paragraph{First attempt: a single predicate.} We define $\alpha$
as follows, for a concrete state~$c$:
\[
  \alpha_1(c) \wbox{$=$} (c.\pc_0, \, c.s = c.l_0) \in \range 0 4 \times
  \{0,1\} \ ,
\]
where $c.\pc_0$, $c.l_0$, and $c.s$ are the values of thread 0's $\pc$ and
local variable $l$, and of shared variable $s$, in state $c$,
respectively. The function extracts from a concrete state the current
program location of thread 0 and the value of predicate\format\linebreak $p
:: (s = l)$ for thread 0.\footnote{Tracking these values for thread 0
  suffices: the multi-threaded program is symmetric.} The only statement
\emph{not} respecting $\alpha_1$ is the \plif\ statement in \lineref{toy:
  if branch}: here, the new value of the $\pc$ cannot be determined from
the current values of $\pc$ and predicate $p$ alone. All other statements
respect $\alpha_1$.

We can now perform an iterative exploration of this program---bounded but
exhaustive within each bound. In \figureref{toy} on the right,
\namecolor{red} arrows denote ``new abstract state reached''. A~red
horizontal arrow (\code{$r$++}) means: ``keep increasing~$r$''. A~red
vertical arrow (\code{$d$++}) means: ``switch to increasing $r$''. In other
words, following a red arrow---no matter the direction---we always go
``right'' (\code{$r$++}). The \namecolor[green]{darkgreen} horizontal arrow
followed by a \namecolor[green]{darkgreen} vertical arrow at the end
indicates that we have reached the first plateaus of length 1 in
\emph{both} directions: at $(r,d) = (7,4)$.

At this point we have reached a total of 7 abstract states. State $(3,0)$
\format(\mbox{$\pc=\lineref[]{toy: assertion}$}, $s \not= l$) is not among
them, so the assertion has not been violated so far. We run the convergence
test, to determine whether set $\absreached[7,4]$ is closed under
disrespectful actions. Since the \plif\ in \lineref{toy: if branch} is the
only disrespectful statement, we only need to check successors of abstract
states of the form $(1,?)$ (i.e., with $\pc=1$). Unfortunately,
$\absreached[7,4]$ contains abstract state $(1,0)$ (a~reachable abstract
state) but not its abstract successor $(2,0)$. This state is unreachable,
but we do not know that at this point. This causes \algorithmref{DUBA} to
return ``unknown''.

\paragraph{Second attempt: two predicates.} The disrespectful action
causing the failure suggests that we need to keep track of whether the
branch in \lineref{toy: if branch} can be taken, i.e.\ whether $m=1$. We
refine our abstraction using this (non-mixed) predicate:
\begin{equation}
  \label{equation: toy: alpha 2}
  \alpha_2(c) \wbox{$=$} (c.\pc_0, \, c.s = c.l_0, \, c.m = 1) \in \range 0 4 \times \{0,1\}^2 \ .
\end{equation}
The abstract successors of the \plif\ statement can now be decided based
only on knowledge provided by $\alpha_2$, i.e.\ the statement respects
$\alpha_2$. There is, however, another statement disrespecting $\alpha_2$,
and only one: the increment \code{$m$++} in \lineref{toy: inc m}. If $m
\not= 1$, we cannot decide whether $m=1$ will be true after the increment.

We again perform our iterative exploration of this program, and find the
first suitable plateau at the same point $(r,d) = (7,4)$. This time,
however, we have reached a total of 12 abstract states (all of them
``safe''). We run the convergence test: we only need to check already
reached abstract states of the form $(0,?,0)$ ($\pc=0$, $m \not= 1$).
% ; from these the \code{$m$++} statement is ambiguous in the abstract.
Set $\absreached[7,4]$ contains exactly one state of this form: $(0,1,0)$,
which \code{$m$++} can turn into $(1,1,0)$ and $(1,1,1)$---note that the
next $\pc$ value is unambiguous (1), and predicate $s = l$ is not
affected. \emphasize{The good news} is now that both abstract states
$(1,1,0)$ and $(1,1,1)$ are contained in $\absreached[7,4]$. This proves
this set closed under disrespectful actions; \algorithmref{DUBA}
terminates: the assertion is safe for any execution schedule, for the case
of $n=2$ threads.

\Paragraph

We summarize that, in our solution above, we assumed a lucky hand in
picking predicates%
%, including the one chosen during refinement---%
%, and one of our predicates was of the mixed type
---the question of predicate discovery is orthogonal to the delay-unbounded
analysis scheme. However, the proof obtained using \algorithmref{DUBA} does
not involve costly broadcast operations, previously proposed as an
ingredient to extend predicate abstraction to concurrent programs. A
second, more powerful advantage is that, unlike the earlier broadcast
solution, \algorithmref{DUBA} extends gracefully to the unbounded-thread
case. This is the topic of the rest of this section.

\subsection{The Unbounded-Thread Case}
\label{section: The Unbounded-Thread Case}

The goal now is to investigate whether an asynchronous unbounded-domain
variable program is safe for \emph{arbitrary} thread counts (and thread
interleavings).

\paragraph{Existing solutions.} We are aware of only one general technique
that combines predicate abstraction with unbounded-thread
concurrency~\cite{KKW17}. That technique can achieve the above goal,
roughly as follows. In addition to standard and mixed predicates used also
in the fixed-thread case, we now permit \emph{inter-thread} predicates,
which quantify over all threads other than the executing one. Such
predicates allow us to express, for example, that a thread's local variable
$l$'s value is larger than that of any other thread: \mbox{$\forall i: i
  \not= \mathit{self}: l > l_i$}. Predicates of this type are provably
required during predicate abstraction to verify the safety of the Ticket
Lock algorithm \cite{A91,KKW17}.

Abstraction against inter-thread predicates leads to a \emph{dual-reference
  program}~\cite{KKW17}, a process that is already far more complex than
standard sequential or even fixed-thread predicate abstraction. But we pay
another price for using these predicates: namely, the loss of
\emph{monotonicity} of the transition relation w.r.t.\ a standard
well-quasiordering $\wqo$ on infinite state sets of unbounded-thread
Boolean programs. In this context, monotonicity states, roughly, that
adding passive threads to a valid transition keeps the transition intact.

This price is heavy, since monotonicity w.r.t.\ $\wqo$ would have given us
a well-quasiordered infinite-state transition system, for which local-state
reachability properties are decidable~\cite{A10}; working implementations
exist. The above-mentioned prior work attempts to salvage the situation, by
adding a set of transitions (the \emph{non-monotone fragment}) to the
dual-reference program that restore monotonicity and further
overapproximate but without affecting the reachability of unsafe
states~\cite{KKW17}.

\paragraph{Alternative solution.} We now propose a solution that uses the
same type of inter-thread predicates (this is inevitable), but renders
dual-reference programs, the monotone closure of the transition relation
and all other ``overhead'' introduced in~\cite{KKW17} unnecessary. We will
use \algorithmref{DUBA} as a sub-routine.

The idea is as follows. \sectionref{The Fixed-Thread Case} suggests a way
to verify fixed-thread asynchronous programs, using a combination of
predicate abstraction and \algorithmref{DUBA}. To handle the
unbounded-thread case, we wrap another layer of incremental resource
bounding around this combined algorithm---the ``resource'' this time is the
number $n$ of threads executing the program. For each member of a sequence
of increasing fixed thread counts we compute the set of abstract states
reachable under arbitrary thread interleavings. This is purely a
sub-routine; we will use the method proposed in \sectionref{The
  Fixed-Thread Case} (others are possible, e.g.~\cite{DKKW11}).

The incremental (in $n$) analysis proceeds until we have reached a thread
plateau \emph{of length 1}, and then run the convergence test: we check the
current abstract reachability set for closure under disrespectful
actions. This time, the abstract transitions must take into account that
the number of executing threads is unknown. It is easy to see that a
plateau of length 1 is sufficient: we compute the set of abstract states
reachable under \emph{arbitrary} thread schedules; thus, the obstacle of
non-schedulability of thread $i$ in the proof of \theoremref{DUBA} that
forced us to wait for a (delay) plateau of length $n-1$ does not apply
here.

\subsection*{A non-monotone resource parameterization}

Before we demonstrate
this idea on program $\otherprg$, we justify our strategy of combining
resource bounds. The idea presented above can be viewed as a multi-resource
analysis problem where we increment $r$ and $d$ in an ``inner loop''
(represented by \algorithmref{DUBA} as a sub-routine to compute
fixed-thread reachability sets), and $n$ in an outer loop. Both loops
compute monotonously increasing reachability sequences: for ``inner'' this
is \propertyref{monotonicity in r and d}; for ``outer'' this is easy to
see. \theoremref{DUBA} relies upon the monotonicity: without it, the test
$\absreached[r,d] = \absreached[r+1,d+n-1]$ makes the algorithm unsound.

The way we nest the three involved resource parameters is not arbitrary:
Round-Robin reachability under an increasing thread count is not
monotone. More precisely, making the thread-count parameter $n$ explicit,
let $\reached[r,d,n]$ denote the set of states reachable in the $n$-thread
program $\prg$ under $\RR[r,d]$ scheduling. Then $\reached[r,d,n] \subseteq
\reached[r,d,n+1]$ is \emphasize{not} valid. The following example
illustrates this (at first counter-intuitive) monotonicity violation:

\

\noindent
\hspace{-4pt}
\centerTwoOut{%
  \begin{minipage}[c]{.48\textwidth}
    \begin{EXA}
      \label{example: monotonicity violation}
      Consider the asynchronous Boolean program over shared variables $s$
      and $t$ on the right. Here we have $\reached[3,0,1] \not\subseteq
      \reached[3,0,2]$: given 1 thread (sequential execution), a state with
    \end{EXA}
  \end{minipage}
}{%
  \fbox{%
    \begin{minipage}[c]{.44\textwidth}
      \renewcommand{\algorithmicrequire}{}
      \begin{algorithmic}[1]
        \setcounter{ALC@line}{-1} % to start numbering with 0
        \REQUIRE{\plkeyword{shared bool} \code{$s$ := 0, $t$ := 0}}
        \STMT \code{$t$ := !$t$} \label{line: non-monotone: flip}
        \IF {$t$}                \label{line: non-monotone: if}
          \STMT \code{$s$ := 1}
        \ENDIF
      \end{algorithmic}
    \end{minipage}
  }
}

\vspace{2pt}

\noindent
{\em $s=1$ is reachable. With 2 symmetric threads, under delay-free
  Round-Robin scheduling ($d=0$), the first and second thread will
  repeatedly flip $t$ to $1$ and back to $0$, resp., before either one has
  a chance to get past the guard in \lineref{non-monotone: if}.

  A stronger result is: for all $r \in \NN$, $\reached[3,0,1] \not\subseteq
  \reached[r,0,2]$, i.e.\ we cannot make up for the poor scheduling of the
  second thread by adding more rounds.\Ifspace{\ This example also
    illustrates the power of the delay-bound parameter.}}

\

\noindent
The consequence for us is that we cannot compute, for fixed $r,d$, the
sets $\reached[r,d,\infty]$, using the closure-under-disrespectful-actions
paradigm. Instead we must, for each $n$, compute
$\reached[\infty,\infty,n]$ (using \algorithmref{DUBA} or otherwise) and
increase $n$ in the outer loop.\Ifspace{\ We remark that \theoremref{DUBA}
  can be fixed to work even for the non-monotone case. However, loss of
  monotonicity in the resource parameter entails all kinds of other
  inconveniences, e.g.\ subset relationships can no longer be determined
  efficiently by comparing set cardinalities.}

\subsection*{Verifying program $\otherprg$ for unbounded thread count}

We recall that, given the two predicates shown in \equationref{toy: alpha
  2} and the pc, we were able to verify program $\otherprg$ correct (under
arbitrary thread interleavings) for $n=2$ threads; a total of 12 abstract
states were reached (out of $5 \cdot 2^2 = 20$ possible). Advancing the
outer loop, we invoke \algorithmref{DUBA} for $n=3$ threads. This reveals
another reachable abstract state, namely $\pc = 0$, $s \not= l$, $m \not=
1$. Unfortunately, this state causes \algorithmref{DUBA} to return
``unknown'': under $\alpha_2$, one currently unreached abstract successor
is $\pc=1$, $s \not= l$, $m = 1$, violating closure. Observing that a
thread executing \lineref{toy: if branch} with $m=1$ must be the first
thread executing, we try tracking the initial value of $m$:
\begin{equation}
  \label{equation: toy: alpha 3}
  \alpha_3(c) \wbox{$=$} (c.\pc_0, \, c.s = c.l_0, \, c.m = 1, \, c.m = 0) \in \range 0 4 \times \{0,1\}^3 \ .
\end{equation}
Interestingly, \emph{all actions (statements) of program $\otherprg$
  respect abstraction $\alpha_3$}. This means that the test for closure
under disrespectful actions is \emph{vacuously true}---we can stop as soon
as we have reached a plateau in $n$ of length 1. We don't have to wait long
for this plateau: we invoke \algorithmref{DUBA} for $n=3$ and $n=4$ under
abstraction~$\alpha_3$. (Note that $n=4$ requires a longer plateau than
$n=3$.) The abstract reachability sets consist of the same 14 abstract
states in both cases. We report the program safe, for arbitrary
interleavings and arbitrary thread counts. We can also report the exact set
of 14 reachable abstract states.

We again summarize that, while we still (and unavoidably) use mixed
predicates, we do not construct a thread-parameterized abstract program,
which would require broadcast statements~\cite{DKKW11} and a rather
involved dual-reference transition semantics~\cite{KKW17}. In fact, we did
not even need to test for closure under any abstract images, since the
chosen abstraction enjoys respect from all actions.

\Ifspace{We note: whenever all actions of a program respect a given
  finite-domain abstraction, the algorithm is guaranteed to succeed: we
  either find a bug, or we reach a plateau---one of these outcomes is bound
  to happen.}

\draftnewpage

\section{Evaluation}
\label{section: Evaluation}

Our goal for the evaluation of \duba\ was to answer the following questions:
\newlist{goals}{enumerate}{10}
\setlist[goals]{label*=\arabic*}
\crefname{goalsi}{goal}{goals}
\Crefname{goalsi}{Goal}{Goals}
\begin{goals}
	\item How does \duba\ compare to abstract fixed-point computation
          (``AI'')\thomasmargin{Suggestion: we say \emph{AI approach} (no
          quotes) but \emph{``AI''} (quotes) when used stand-alone:
          \emph{faster than ``AI''}. Similarly
          \emph{CUBA}/\emph{''CUBA''}}? \label{AI}

	\item How does \duba\ compare to the approach from~\cite{LW18} (``CUBA'')?
%, on the same benchmarks?
          \label{CUBA}

	\item How expensive is the state exploration along a plateau in \algorithmref{DUBA}? \label{Plateau}

	\item What is the performance benefit of the frontier optimization in \algorithmref{DUBA}? \label{Frontier}
\end{goals}
Questions 1 and 2 serve to compare \duba\ against other techniques;
Questions 3 and 4 investigate features of \algorithmref{DUBA}.

%\subsection{Experimental Setup}

To this end we implemented, in Java 11, a verifier using
\algorithmref{DUBA} that takes
% on the same language of 
concurrent pushdown systems as input; we refer to this verifier as
\duba\ in this section.\footnote{DrUBA implementation available at \url{https://doi.org/10.5281/zenodo.4726302}}
%used in~\cite{LW18}. \\
We also implemented the AI approach in Java 11. For the comparison with the
context-unbounded approach, we used a publicly available
tool\footnote{\url{https://github.com/lpzun/cuba}}. Our experiments are
based on the concurrent benchmark programs also used in~\cite{LW18}.
%, to effectively compare with the CUBA runtimes.
The experiments are performed on a 3.20GHz Intel i5 PC. The memory limit
was 8GB, with a timeout of 1 hour.

%% \duba\ takes a CPDS as input.
%% Given a CPDS, a state is represented as the global state and the contents
%% of each stack, along with scheduling information. An abstract state is the
%% global state and the top of each stack.
\subsection{Results}
\tableref{AcrossApproaches} reports the benchmark names, the thread counts,
and the size of the reachable abstract state space (columns 1--3). The
second part of the table shows the time it took each verifier to fully
explore the state space and confirm convergence. For the AI approach, we
check whether the abstract state space is closed under \emph{all}
operations each time either $r$ or $d$ is
incremented. \algorithmref{DUBA}\andrewmargin{note to discuss lines in
  table} was faster than ``AI'' on every example except Stefan-4,5. Stefan
is the only program that actually does not require any delays to discover
all reachable abstract states. The results indicate that the AI approach
spends approximately half of its computation time doing repeated
convergence tests after each bound increment. Furthermore, as state sets
increase in size, AI seems to take even longer, as with the Bluetooth3
(2+3) example. The convergence test needed for ``AI'' includes checking
closure under both respectful and disrespectful actions, making it more
costly than the one used in \algorithmref{DUBA}.
% This answers Question~\ref{AI}.
\begin{table}[htbp]
  \caption{Benchmark description and running times for different
    algorithms. Threads: \# of threads ($a+b$: the respective numbers of
    threads from two different templates); $\absreached$: number of
    reachable abstract states; Time: running time (sec) for each algorithm
    ("---": timeout or memory-out); $r_{\max},d_{\max}$: round and delay
    counts at the \emph{end} of each plateau when convergence was
    detected.}
  \label{table: AcrossApproaches}
  \small
  \begin{center}
    \begin{tabular}{r|lc|r||r|r|r|c|}
      \mc{1}{c}{} & \ch 1 & \ch 2 & \ch 3 & \ch 4 & \ch 5 & \ch 6 & \ch 7 \\
      \cline{2-8}
        &&&&&&&\\[-1em]
        & Benchmark & Threads & \mc 1 {c||} {$|\absreached|$} & \Time               & \Time           & \Time & $r_{\max},d_{\max}$ \\
        &           &         &                               & \algorithmref{DUBA} & \mc 1 {c|} {``AI''} & ``CUBA''  & \algorithmref{DUBA}  \\
      \cline{2-8}
        &		&1+1			&1010			& .69				& .92 			& .32  		&  23, 15 \\
      1 & Bluetooth1		&1+2			&5468			& 3.32			& 6.79 		& 2.25  		&  32, 29 \\
	&		&2+1			&18972		& 8.52			& 16.69 		& 13.60  		&  33, 26 \\
      \cline{2-8}
        &		&1+1			&1018			& .71				& .98			& .29  		&  23, 15 \\
      2 & Bluetooth2		&1+2			&5468			& 3.60			& 6.68 		& 2.62		&  32, 29 \\
        &		&2+1			&18972		& 8.81			& 16.68		& 13.97  		&  33, 26 \\
      \cline{2-8}
        &		&1+1			&1018			& .72				& 1.23		& .41  		&  23, 15\\
        &		&1+2			&5468			& 3.61			& 6.40 		& 2.79  		&  32, 29 \\
      3 & Bluetooth3		&2+1			&19002		& 9.97			& 16.27		& 14.50  		&  33, 26 \\
        &		&2+2			&94335		& 70.71			& 136.31		& 343.05  		&  44, 40 \\
        &		&2+3			&460684		& 654.47			& 2084.76		& TO  			&  56, 56 \\
      \cline{2-8}
        &		&1+1			&272			& .36				& .49			& .14  		&  31, 16 \\
      4 & BST-Insert		&2+1			&6644			& 3.62			& 5.25 		& 10.09  		&  49, 32 \\
        &		&2+2			&14256		& 8.12			& 14.87		& 99.94  		&  50, 38 \\
      \cline{2-8}
      5 & Filecrawler		&1+2			&246			& .37				& .54			& .05 		&  20, 12 \\
      \cline{2-8}
      6 & K-Induction	&1+1			&130			& .51				& .74 			& .48  		&  20, 09 \\
      \cline{2-8}
      7 & Proc-2		&2+2			&130			& .56				& .77 			& 2.05 			&  19, 20 \\
      \cline{2-8}
        &		&2			&31			& .24				& .36 			& .04  		&  13, 02 \\
      8 & Stefan		&4			&687			& 13.99			& 13.82		& 20.33 		&  32, 04 \\
	&			&5			&3085			&428.22			&295.02		& OOM & 35, 05\\
        &		&8			& ---			& OOM				& OOM			& OOM  			& --- \\
      \cline{2-8}
      9 & Dekker		&2			&1507			&      .82				& 1.62 		& .39  		&      37, 16 \\
      \cline{2-8}
    \end{tabular}
  \end{center}
\end{table}

\algorithmref{DUBA} also improved on the results with ``CUBA''. For
examples that took longer than a few seconds, \duba\ was able to run in
less time on the same benchmark. The difference on small examples is likely
due to a different implementation language (C++ vs.\ Java). \duba\ does not
explore as many schedules, and explores fewer as the delay and round bounds
approach their cutoff values (as noted below). Additionally, \duba\ was
less memory-intensive for large examples for which the CUBA approach cannot
prove that the set of reachable states per context bound is finite. In this
case, ``CUBA'' requires the use of more expensive symbolic representations
of states sets. \algorithmref{DUBA} does not suffer from this problem---the
reachability sets in each iteration are finite. For the Stefan-5 example,
``CUBA'' ran out of memory after 23 minutes. \duba\ was able to prove
convergence for this example (as was ``AI'').
% This answers Question ~\ref{CUBA}.
%% , as our approach executes the concrete program for a finite number of
%% steps, regardless of the properties of the program.

\tableref{WithinDrUBA} reports the number of times \algorithmref{DUBA}
computed the image (successors) of a state until reaching the final
$r$-$d$-plateau (Col.~3) and during the final plateau (Col.~4), as well as
the total number of image computations without the \emph{frontier}
optimization (Col.~5). The table offers convincing evidence to support our
heuristic that waiting for a long $d$-plateau at the end of exploration is
not costly, answering Question~\ref{Plateau}. On most benchmarks, the
amount of computation done during the plateau (Col.~4) was negligible. This
included our largest example, Bluetooth3 (2+3). The exception to this is
the Stefan examples, which---as mentioned earlier---do not require any
delays to reach the full abstract state set (the $d$-plateau starts at
($r_{\max}$,0)). Finally, a naive implementation that does not take
advantage of monotonicity, forgoing the frontier approach to expanding the
state set, was orders of magnitude worse. This is because it has to
recompute the whole set for every iteration of $r$ or $d$. This answers
Question~\ref{Frontier}.
\begin{table}[htbp]
  \caption{Detailed analysis of \algorithmref{DUBA}, measuring the number
    of times the program computed the successors of a state. Col.~3 reports
    the image operations \algorithmref{DUBA} performed before reaching the
    FP (final plateau), Col.~4---the number of additional image operations
    computed until the program ended. Col.~5 shows the image operations
    without the \emph{frontier} improvement, requiring recomputing each
    $\absreached(r,d)$ from the initial states.}
  \label{table: WithinDrUBA}
  \small
  \begin{center}
    \begin{tabular}[h]{r|lc||r|r|r|}
      \mc{1}{c}{} & \ch 1 & \ch 2 & \ch 3 & \ch 4 & \ch 5 \\
      \cline{2-6}
      &&&&&\\[-1em]
      & Benchmark 		&Threads		& calls to $\Image$			& calls to $\Image$			& calls to $\Image$       \\
      &        		&			& $\rightarrow$ begin FP 	& begin $\rightarrow$ end FP 	& w/o frontier \\
      \cline{2-6}
      &			&1+1			& 4,034						& 1						& 339,261 \\
      1 & Bluetooth1        &1+2			& 23,441					& 3						& 4,758,084 \\
      &		&2+1			& 80,283					& 19						& 23,199,458\\
      \cline{2-6}
      &			&1+1			& 4,103						& 1						& 350,587\\
      2 & Bluetooth2	&1+2			& 23,493					& 3						& 4,780,778\\
      &		&2+1			& 80,714					& 19						& 23,290,556\\
      \cline{2-6}
      &			&1+1			& 4,096						& 8						& 348,851\\
      &			&1+2			& 23,496					& 3						& 4,786,950\\
      3 & Bluetooth3	&2+1			& 80,834					& 19						& 23,467,470\\
      &		&2+2			& 478,426					& 2						& 283,910,446\\
      &		&2+3			& 2,766,625					& 6						& ---\\
      \cline{2-6}
      &			&1+1			&780						& 1						& 82,130\\
      4 & BST-Insert	&2+1			& 29,802					& 6						& 17,785,065\\
      &		&2+2			& 62,190					& 25						& 34,335,106\\
      \cline{2-6}
      5 & Filecrawler	&1+2			& 1,056						& 4						& 202,074\\
      \cline{2-6}
      6 & K-Induction	&1+1			& 5,636						& 974						& 218,715\\
      \cline{2-6}
      7 & Proc-2		&2+2			& 2,501						& 1,298					& 578,099\\
      \cline{2-6}
      &		&2			&367						& 59						& 500,494\\
      8 & Stefan		&4			& 658,696					& 261,881					& ---\\
      &		&5		& 10,299,275					& 16,920,446					&---\\
      &		&8			& ---						& ---						& ---\\
      \cline{2-6}
      9 & Dekker		&2			& 3,636						& 2						& 688,836\\
      \cline{2-6}
    \end{tabular}
  \end{center}
\end{table}

Comparing Col.~7 in \tableref{AcrossApproaches} to the cutoff
context-switch bounds from~\cite{LW18}, we find that, while the $r$ and $d$
bounds were large, not all programs that needed large bounds took a long
time to verify. For example, the Bluetooth3 (2+1) example took much less time
than Stefan-5, despite requiring 21 more delays (with similar rounds). A
hint for the reason can be found in \tableref{WithinDrUBA}. Once the set
of abstract states is close to the $\absreached$, there are very few new
states on the frontier. We can see this in the small numbers in Col.~4, but
it also applies to the round bound. If a state is rediscovered, it is not
expanded in further round increments. Once the round bound is large enough,
there are few deep schedules of maximum possible length ($nr$)
that produce new concrete states.

\subsection{Unbounded-Thread Experiments}
\label{section: Unbounded-Thread Experiments}

We implemented \algorithmref{DUBA} in combination with predicate
abstraction as detailed in \sectionref{The Unbounded-Thread Case} to check
the effectiveness of our technique on a tricky concurrent program that
requires unbounded variable domains. The Ticket Lock protocol~\cite{A91}
and the predicates used to prove its correctness are shown in
\figureref{ticket lock}. In \lineref{while s not l}, threads wait to enter
the Critical Section, whose code is at the beginning of \lineref{critical
  section}; the rest of \lineref{critical section} is exit code to prepare
the thread for re-entry. In the predicates on the right, subscript $i$
denotes thread $i$'s copy of a local variable.
\begin{figure}[htbp]
  \centerTwoIn{%
    \fbox{%
      \begin{minipage}{.5\textwidth}
        \renewcommand{\algorithmicrequire}{}
        \begin{algorithmic}[1]
          \setcounter{ALC@line}{-1} % to start numbering with 0
          \REQUIRE{\plkeyword{shared int} \code{$s$ := 0, $t$ := 0}}
          \REQUIRE{\plkeyword{local  int} \code{$l$ := fetch\_and\_add($t$)}}
          \STMT\plkeyword{while} \code{$s \not= l$} \plkeyword{do}\code ; \COMMENT{wait for $s=l$} \label{line: while s not l}
          \STMT \emph{critical-section code here} \\
          \code{inc($s$)}   \label{line: critical section} \\
          \code{$l$ := fetch\_and\_add($t$)} \\
          \plgoto\ 0
        \end{algorithmic}
      \end{minipage}}}{%
    \begin{minipage}{.3\textwidth}
      \begin{description}

      \item[P1:] $\forall i: t > l_i$

      \item[P2:] $|\{i: \pc_i = 1\}| \atl 2$

      \item[P3:] $s = l$

      \item[P4:] $\forall i: i \not=\mathit{self}: l \not= l_i$
      \end{description}
    \end{minipage}}
  \caption{\emphasize{Left:} the Ticket Lock protocol; \emphasize{Right:}
    four predicates used to prove it correct}
  \label{figure: ticket lock}
\end{figure}
  %% \begin{minipage}{.45\textwidth}
  %%   \hspace*{1mm}\textbf{P1:} $\forall i: t > l_i$\\
  %%   \hspace*{1mm}\textbf{P2:} (\# of threads with $pc = 1) \geq 2$\\
  %%   \hspace*{1mm}\textbf{P3:} $s = l$\\
  %%   \hspace*{1mm}\textbf{P4:} $\forall i \not=$ self: $l \not= l_i$
  %% \end{minipage}

This example has been shown to require significant adjustments to predicate
abstraction to accommodate fixed-thread concurrency~\cite{DKKW11}, and has
been claimed to require an entirely new theory to cope with the
unbounded-thread case~\cite{KKW17}. We rely on the same predicates used in
earlier work, and it is clear what motivates each predicate. P1 ensures $t$
is a "new" ticket larger than previous ones, P2 is used to check the safety
property, P3 tracks the condition in Line~0, and P4 means that the
operating thread's $l$ is unique. \duba finds four abstract states for both
2 threads and 3 threads using \algorithmref{DUBA}. This is an $n$-plateau
of length 1.

To prove convergence for both \algorithmref{DUBA} and the ``outer loop''
incrementing~$n$, we used the ACL2s theorem prover
\cite{DBLP:journals/entcs/DillingerMVM07}. We specified the data in a
concrete state, and the four abstract states that were found. Only the
second statement disrespects this abstraction w.r.t. $r,d$ and $n$, as we
know the value of the test in the first statement for an abstract
state. Given these, ACL2s was able to verify that the set of abstract
states is closed under the semantics of statement 1. As a result, we can
report that Ticket Lock is safe (P2 is invariantly false), for an arbitrary
number of threads and arbitrary thread interleavings.

\draftnewpage

\section{Discussion of Related Work}
\label{section: Discussion of Related Work}

This work is inspired from two angles. The first is clearly the
delay-bounded scheduling (DBS)
technique~\cite{DBLP:conf/popl/EmmiQR11}. The authors formalize this
concept and show its effectiveness as a testing scheme. Their computational
model of a dynamic task buffer is somewhat different from ours. We have not
discussed dynamic thread creation here; it can be simulated by creating
threads up-front and delaying them until such time as they are supposed to
come into existence.\andrew{\ If we make our RR scheduler wait to add
  machines for scheduling until the next round, it is fine. Worst case here
  is a system where every machine makes a "dummy" machine that undoes the
  interesting interaction between 2 machines. In this case, we need a huge
  (but finite) number of delays to get any interesting interaction between
  the 2 machines (something like $r^2n$ where n is the initial/interesting
  machine count)} The DBS paper also presents a sequentialization technique
that can be turned into a symbolic verifier via verification-condition
generation and SMT solving. This, however, requires bounding loops and
recursion. Our approach combines exhaustive finite-state model exploration
with convergence detection and thus does not suffer from these
restrictions.

The second inspiration comes from an earlier context-unbounded analysis
technique~\cite{LW18}. Similar in spirit to the present work,~\cite{LW18}
started from a yet earlier context-bounded analysis technique and describes
a condition under which a chosen context bound is sufficient to reach all
states reachable under some abstraction. For the case of concurrent
pushdown systems (CPDS)---the verification target of~\cite{LW18}---, the
pop operation plays a crucial role in establishing this condition; note
that, in our work, pop actions disrespect the top-of-the-stack abstraction
commonly used for CPDS.

Our work has a number of advantages over~\cite{LW18}. First, and crucially,
the set of states reachable under a context bound can be infinite (a single
context can already generate infinitely many states); its determination
thus requires more expensive symbolic reachability methods.
%%  (the paper offers a
%% sufficient but not necessary condition for the finiteness of all
%% context-bounded reachability sets).
In contrast, the reachability set under Round-Robin scheduling with a
round- and a delay bound \emph{is always finite}; moreover, it can be
computed very easily, even for complex programs. This makes our technique a
prime choice for lifting existing testing schemes to verifiers. A second
advantage over~\cite{LW18} is that we retain much of the efficiency of the
``almost deterministic'' exploration delay-bounded scheduling, as
demonstrated in \sectionref{Evaluation}. A downside of our work is that our
convergence condition is sound only after a plateau has emerged of length
roughly equal to the number of running threads; this is not required
in~\cite{LW18}. However, as also demonstrated in \sectionref{Evaluation},
our efforts to compute reachable states for increasing $r$, $d$ in a
\emph{frontier-driven} way nearly annihilates this drawback: in most cases,
only a small number of image computations happen along the plateau.

An alternative to our verification approach is a classical analysis based
on abstract interpretation~\cite{CC77}. Given function $\alpha$, such
analysis interprets the entire program abstractly, and then computes a
fixed point under the abstract program's transition relation. This fixed
point, if it exists, overapproximates the set of reachable abstract
states. Hence, the absence of error states in the fixed point implies
safety, but the presence of errors does not immediately permit a
conclusion. In contrast, our technique interleaves \emph{concrete} state
space exploration (enabling genuine testing) with \emph{abstraction-based}
convergence detection. We believe this to be a useful approach in practical
programming environments, where abstract proof engines with poorly
understood bug-finding capabilities may be met with skepticism. A more
detailed discussion of \duba\ vs.~Abstract Interpretation can be found in
\Ifthenelse{\arxiv}{\appendixref{Delay-Unbounded Analysis vs. Abstract
    Interpretation}}{the Appendix of~\cite{JW21a}}\thomasmargin{\version}.

Underapproximating program behaviors using bounding techniques is a
wide-spread solution to address undecidability of safety verification
problems. Examples include depth- \cite{G97} and context-bounding
\cite{QR05,LMP09,LR09}, delay-bounding~\cite{DBLP:conf/popl/EmmiQR11},
bounded asynchrony~\cite{FHMP08}, preemption-bounding~\cite{MQ07}, and
phase-bounded analysis \cite{BE14,AAC13}. Many of these bounding techniques
admit decidable analysis problems \cite{QR05,LMP09,LR09} and thus have been
successfully used in practice for bug finding. Round- and delay-bounded
Round-Robin scheduling trivially renders safety decidable, since the
delay-program is finite-state. In addition, it is very easy to implement,
avoiding, for example, the need for symbolic data structures and algorithms
to represent and process intermediate reachability sets.

\section{Conclusion}
\label{section: Conclusion}

We have presented an approach to enhancing delay-bounded scheduling in
asynchronous programs with a convergence test that, if successful,
certifies that all states from some chosen abstract domain have been
reached. The resulting algorithm inherits from earlier work the capability
to detect bugs efficiently, but can also prove safety properties, under
arbitrary thread interleavings. It exploits the monotonicity
% and finiteness
of delay-bounded reachability sets to expand states and test for
convergence only when needed. We have further demonstrated that, combined
with predicate abstraction using powerful predicates, tricky
unbounded-thread routines over unbounded data, such as the Ticket Lock, can
be verified using substantially less machinery than proposed in earlier
work. We have shown the experimental competitiveness of our approach
against several related techniques.

\clearpage

\appendix

\section*{Appendix}

\section{Proofs of Claims}
\label{appendix: Proofs of Claims}

We include here the proofs of all claims made formally in this paper that
were omitted earlier. The numbers of the following theorems, lemmas, etc.,
are identical to their respective numbers in the main text.

\setcounter{ASS}{\theDUBAisFREE}
\begin{THE}
  State $s$ is reachable under free scheduling iff there exist $r,d$ such
  that $s$ is reachable under $\RR[r,d]$ scheduling: $\reached =
  \Union_{r,d \in \NN} \reached[r,d]$.
\end{THE}
\Proof: the $\Leftarrow$ direction requires no further proof, since a free
scheduler permits any schedule. For the other direction, consider a path $p
= (\range[]{s_0}{s_l})$ to $s=s_l$. We have to show that there exist $r$,
$d$ such that the conditions in \definitionref{reachable under RR with r
  and d} hold. Path $p$ remains the same. We have to provide the function
$f$ with the required properties, for suitable $r$ and $d$.

To this end, let $q = (\range[]{q_0}{q_{l-1}})$ be the sequence of threads
executing in each of the $l$ steps along $p$, and let $\func f {\NN}
{\range{0}{n-1}}$ be the function defined by $f(i) = q_i$ for $0 \atm i <
l$ and $f(i) = 0$ otherwise. Let further $d_p$ be the total delay (defined
in item~\itemref[]{d_p} of \definitionref{reachable under RR with r and
  d}). Now we simply set $d := d_p$ and $r := \ceils{(l + d_p) / n}$. This
immediately satisfies conditions \itemref[]{d_p}.\ and \itemref[]{r
  bound}. Condition \itemref[]{scheduling constraint} holds by the
definitions of $q$ and $f$.\eop

\setcounter{ASS}{\thePlateauCondition}
\begin{LEM}
  If \algorithmref{DUBA} reaches \lineref{convergence test}, the current
  values of $r$ and $d$ satisfy: \\
  (i)~$\Reached = \reached[r,d]$, and (ii) $\absreached[r-1,d-(n-1)] =
  \absreached[r,d]$.
\end{LEM}
\Proof: consider property (i);
% $\Reached = \reached[r,d]$ when the algorithm exits the delay loop.
we show $\Reached \subseteq \reached[r,d]$ and $\Reached \superseteq
\reached[r,d]$ separately.

We first show that ``$\subseteq$'' is a loop invariant for both repeat
loops. This subset relationship holds after initializing $r$ and $d$ in
\lineref{initialize r,d}, at which point $\Reached$ contains exactly the
initial states, a subset of $\reached[0,0]$.

For the round loop beginning in \lineref{round loop begin}, states are only
added to $\Reached$ in \lineref{for loop round body}. Here, the value of
$r$ is one greater than the values for all of the states being expanded,
since they were chosen with $rounds\_taken = r$ and then $r$ was
incremented. When \algorithmref{FinishRounds} is called with these states,
it will not expand past the given round bound. The delays of the states are
not changed. Since this is the only addition of states to $\Reached$ in the
loop, no state added here can take more than the current $r,d$ in
\lineref{for loop round body}, and the values of $r$ and $d$ are not
changed before line \lineref{round loop end}, this invariant
holds. Finally, We need to show that when we re-enter this loop from
\lineref{start over}, $\Reached \subseteq \reached[r,d]$ is still
true. This is implied as it is also a loop invariant for the delay loop. So
$\Reached \subseteq \reached[r,d]$ is a loop invariant for the round loop.

For the delay loop beginning in line \lineref{delay loop begin}, we know
that $\Reached \subseteq \reached[r,d]$ is true because we just left the
round loop. Next, states are only added in \lineref{finishrounds
  delay}. The states \algorithmref{FinishRounds} is called on will have
delays equal to $d$, since they were chosen with maximum delays, $d$ was
incremented, and then they were delayed one time. Since
\algorithmref{FinishRounds} does not delay states, this will remain true
for all of the states added. \algorithmref{FinishRounds} also will not
expand past $r$ rounds, and since all of the states it is called on started
with at most $r$ rounds, this will remain true for all states added. So,
$\Reached \subseteq \reached[r,d]$ is a loop invariant for the delay
loop. It is also the true in \lineref{start over}, for the same reasons, so
it is true whenever the algorithm enters the round loop.

Next, we show that $\reached[r,d] \subseteq \Reached$ when we exit the
delay loop after \lineref{delay loop end}. Consider a path from $i$ to a
state $c \in \reached[r,d]$:
$$i \stackrel{\mboxscript{$0,x_0$}}{\longrightarrow} i'
\stackrel{\mboxscript{$1,x_1$}}{\longrightarrow}
... \stackrel{\mboxscript{$t_{d-1},x_{d-1}$}}{\longrightarrow} d
\stackrel{\mboxscript{$t_{d},delay$}}{\longrightarrow} d'
\stackrel{\mboxscript{$t_{d+1},x_{d+1}$}}{\longrightarrow}
... \stackrel{\mboxscript{$t-1,x_{t-1}$}}{\longrightarrow} b
\stackrel{\mboxscript{$t,x_t$}}{\longrightarrow} c$$ Above, the path starts
at an initial state and ends in some state $b$ which is taken by thread $t$
to $c$ using transition $x_t$. We have also shown the last delay in the
path, which is just represented by adding a transition "delay" which only
changes the scheduling variables, taking some state $d \to d'$ in the path
and delaying thread $t_d$.\\ Every such path is possible up to $r$ rounds
with $d$ delays. Each path is made up of two types of transitions, normal
transitions and delays. We can show the path is possible in the algorithm
by induction on the number of delays in the path. First, every path with 0
delays up to the round bound is possible. These are found immediately in
the \code{repeat} loop from Lines~\lineref[]{round loop
  begin}--\lineref[]{round loop end}. Now, consider a path with $d$
delays. As shown above, we can split the path into a path with $d-1$ delays
from state $i \to d$, the last delay, and a path with 0 delays from $d' \to
c$. The path from $i \to d$ is possible by the inductive hypothesis. Next,
the delay is possible because $d$ will be included in the $Frontier$ in
\lineref{delay frontier}, and so will get delayed in the for loop and
delayed once. Finally, the last part of the path with 0 delays will be
possible from the $\FinishRounds$ call in line \lineref{finishrounds
  delay}. So, after the \code{for} loop in line \lineref{delay increment},
all schedules with $r$ rounds and $d$ delays will have been possible. If
the delay plateau condition is met in line \lineref{check delayed state
  new}, then $\Reached$, $r$, and $d$ are not changed before we exit the
delay loop, so this remains true.

Hence, after exiting the delay loop, we have $\Reached \subseteq
\reached[r,d]$ and $\reached[r,d] \subseteq \Reached$, thus $\reached[r,d]
= \Reached$. The algorithm does not change these before it returns with
either answer.

\

Property (ii) follows directly from the fact that, upon exit of the second
\plrepeat\ loop, neither loop has generated any new abstract states
(otherwise \lineref{start over} would have caused a return to the beginning
of the first loop in \lineref{round loop begin}). Moreover, by the loop
exit conditions, the plateaus have length $1$ and $n-1$ in $r$ and~$d$,
resp.\eop

\setcounter{ASS}{\theDUBAsoundness}
\begin{COR}
  The answers ``safe'' and ``violation of $\prop$'' returned by
  \algorithmref{DUBA} are correct.
\end{COR}
\Proof: \algorithmref{DUBA} returns ``safe'' only in \lineref{safe}. By
\lemmaref{plateau condition}, the test performed in \lineref{convergence
  test}, and \theoremref{DUBA}, we have $\absreached[r,d] =
\absreached$. Suppose $c$ was a reachable concrete state that violates
$\prop$. Let $c'$ be any concrete state already reached such that
$\alpha(c') = \alpha(c)$ (we know that $c'$ exists since all reachable
abstract states have been reached). Since $\prop$ respects $\alpha$ (see
beginning of \sectionref{Efficient Delay-Unbounded Analysis}), $c'$
violates $\prop$ as well. But no violation of property $\prop$ has been
flagged so far (the program would not have reached \lineref{convergence
  test}). So states $c'$ and $c$ do not exist; the program is safe.

\algorithmref{DUBA} returns ``violation of $\prop$'' only in
\lineref{violation} of \algorithmref{FinishRounds}. At that point, state
$u$ has been found to violate $\prop$. State $u$ is reachable, since states
added to $\Unexplored$ (and to $\Reached$) are generated only by the
$\Image$ function, which naturally maps reachable to reachable
states.\eop

\setcounter{ASS}{\theDUBAtermination}
\begin{LEM}
  If the domain $\abstrStates$ of abstraction function $\alpha$ is finite,
  \algorithmref{DUBA} terminates on every input.
\end{LEM}
\Proof: independently of the finiteness of $\abstrStates$, we first argue
that, at any time, the set $\Reached$ contains only finitely many states:
this is true at the beginning (\lineref{initial states}), due to the
finiteness of the initial-states set. Set $\Reached$ increases after calls
to function $\FinishRounds$ (\algorithmref{FinishRounds}). This function
terminates, returning a finite number of states, since it explores states
Round-Robin style, with the given round bound $r$. Exploration ends with
states encountered during round $r$ by thread $n-1$.

We now discuss the two \plrepeat\ loops. Assuming no ``violation of
$\prop$'' is reported (which would imply termination), the algorithm
reaches \lineref{convergence test} (which implies termination) if the
plateaus in $r$ and $d$ have reached length $1$ and $n-1$, resp. If the
domain $\abstrStates$ of $\alpha$ is finite, then so is set $\absreached$,
hence at some point the abstract reachability sets (which determine whether
we have reached a plateau) will stabilize, i.e.~remain constant henceforth,
and arbitrarily long plateaus will eventually materialize.\eop

\section{Predicate-Abstraction Primer}
\label{appendix: Predicate-Abstraction Primer}

A short primer on predicate abstraction is given in \figureref{predicate
  abstraction primer}.
\begin{figure}[htbp]
  \centerOne{%
    \fbox{%
      \begin{minipage}[c]{.9\textwidth}
        \footnotesize
        Consider $z$ predicates $\range[]{p_1}{p_z}$ on global states $s$ and a
        function $\alpha$ defined by $\alpha(s) =
        (\range[]{p_1(s)}{p_z(s)})$, mapping the state space into the
        finite set~$\{0,1\}^z$. \emph{Predicate abstraction}
        \cite{GS97,BMMR01} refers to the process of statically translating,
        statement by statement, a source program $\otherprg$, typically
        over variables with unbounded domains like integers, to a Boolean
        program $\Bprg$ over variables with domain $\{0,1\}$, one variable
        $b_i$ per predicate $p_i$. The idea is that each statement in
        $\Bprg$ approximates how the corresponding statement in $\otherprg$
        affects the truth of each predicate. The translation is constructed
        such that, for any state $s$ reachable in $\otherprg$, the abstract
        state $\alpha(s)$ is reachable in $\Bprg$. As a result, $\Bprg$ can
        be used to prove safety properties for~$\otherprg$. Like with all
        conservative abstractions, errors detected in $\Bprg$ require
        spuriousness analysis and possibly abstraction refinement before
        conclusions can be drawn.
      \end{minipage}
    }
  }
  \caption{A predicate abstraction primer}
  \label{figure: predicate abstraction primer}
\end{figure}

\section{Delay-Unbounded Analysis vs.~Abstract Interpretation}
\label{appendix: Delay-Unbounded Analysis vs. Abstract Interpretation}

While our proposed technique explores the reachability states in the
concrete domain and uses abstraction only to detect convergence of some
``view'' of the concrete reachability set, classical approaches based on
Abstract Interpretation (``AI'')~\cite{CC77} perform abstract fixed-point
computation: they explore abstract reachability sets, never computing their
full concrete counterparts, weak-until a fixed-point is reached. In
principal terms, our approach can both prove and refute, while AI alone is
not able to demonstrate the existence of bugs.

Our approach pays for this ability with the potentially high cost (in terms
of time and memory) of concrete-state exploration. The availability of an
efficient concrete state-space explorer is therefore essential. These are
often implemented in high-performing testers, which may employ
memory-saving ``tricks'' like exact state-hashing. If applicable to the
program at hand, symmetry reduction is another highly effective
state-saving tool.

\Paragraph

In the following we discuss some details of why---or when---we believe our
approach to be advantageous to verification using AI.

\paragraph{Cost of abstract image computation.} Instead of having to define
(and implement) the full abstract transformer (where mistakes jeopardize
soundness), using our technique we determine abstract successors only for a
subset of the actions, namely for disrespectful ones. For example, for
concurrent threads running recursive procedures, only procedure returns
(``pops'') are dis\-re\-spect\-ful---we do not need to compute abstract
images for any other statements in the program, including calls and
procedure-local ones.\footnote{Prior work has shown how to compute abstract
  successors under procedure call returns efficiently in concurrent
  recursive Boolean programs~\cite{LWR20}.}

Consider the special case that \emph{all} actions respect $\alpha$: by
definition, then, every set is closed under disrespectful actions, and we
can stop the exploration as soon as a plateau of sufficient length is
found---no abstract successors ever need to be computed. An example is
given in \sectionref{The Unbounded-Thread Case}.

\paragraph{Identifying respectful actions.} To unleash the full potential of our
technique, function $\alpha$ must permit some program actions to be
provably respectful. We have shown in \exampleref{DUBA in CPDS} that this
is the case for concurrent pushdown systems.

Generalizing from this case, respectful actions are easily identifiable
whenever the abstraction is a \emph{projection}: suppose a concrete state
$c$ can be written as a pair $c = (a,h)$ (``abstract'' and ``hidden''
part), such that the abstraction simply suppresses $h$: $\alpha(c) =
a$. This applies to the common case of abstractions that project some
unbounded data structure of the program (such as a stack or a queue) to
some (finite) part of it (such as the top of the stack or the head of the
queue) (\exampleref{DUBA in CPDS}; \cite{LWL19,LWR20}).

Projections allow the abstract domain to be structurally embedded into the
concrete domain. This means that the abstract successor function, which
operates on the $a$ part, can be thought of as acting on a concrete state
$c$ where the $h$ part happens to be unknown or nondeterministic. We can
now decide whether some action $x$ respects $\alpha$: namely, when the
abstract parts of the successor states of $c=(a,h)$ do not depend on
$h$. This guarantees that states equivalent under $\alpha$ (i.e., with the
same abstract part) give rise to equivalent successors. Consider sequential
pushdown systems: we split a state into the abstract part ``the top of the
stack'', and the hidden part ``rest of the stack''. The abstract parts of
successors under overwrite or push actions depend only on the current top
of the stack, i.e.\ the abstract part.

\paragraph{Benefit of laziness.} The convergence test is clearly expensive,
so avoiding it in all but certain special $(r,d)$-iterations saves
resources. This can be particularly beneficial if the program is buggy: we
may catch the bug before doing any convergence checking at all. This is of
utmost value, since every convergence check is obviously bound to fail if
the program is buggy and the bug has not been reached yet.

If the abstraction $\alpha$ is precise enough that it does not permit
\emph{intermediate} plateaus (see discussion after \lemmaref{DUBA
  termination}), we are guaranteed to find any error before ever checking
for convergence. In practice, intermediate plateaus appear to be rare, and
we benefit from their absence even if this absence cannot be proved a
priori.

\paragraph{Cost of laziness.} Delaying convergence checking while waiting
for a plateau means we compute more reachable states, for increasing values
of $r$ and~$d$. This potential cost motivates our strictly incremental,
frontier-based reachability computation.

\Paragraph

An approach somewhere between abstract interpretation and our technique is
to perform incremental concrete state-space exploration, and check for
abstract convergence regularly, for instance every time reachability is
complete for the current parameter values. Our empirical evaluation
presented in \sectionref{Evaluation} shows the benefits of ``delaying'' the
convergence check until a plateau of sufficient length has emerged. We
attribute these benefits to two circumstances: the frontier-driven
incremental exploration mentioned earlier, and the fact that our abstract
images are ``smaller'': we only need to consider disrespectful actions,
instead of having to define (and implement) the full abstract transformer,
for all program statements.

\end{document}